%% file: main.tex
\documentclass{WileyMSP-template} 

\usepackage[T1]{fontenc}          
\usepackage[utf8]{inputenc}       
\usepackage{lmodern}              
\usepackage{amsmath, amssymb, bm} 
\usepackage[version=3]{mhchem}    
\usepackage{siunitx}              
\usepackage{graphicx}             
\usepackage{booktabs}             
\usepackage{array, tabularx}      
\usepackage{makecell, multirow}   
\usepackage{upgreek}              
\usepackage{gensymb}              
\usepackage{xspace}               
\usepackage{chemformula}          
\usepackage{nicefrac}             
\usepackage{qcircuit}             
\usepackage{enumitem}             
\usepackage{ragged2e}             
\usepackage{float}                
\usepackage{color,xcolor}         
\usepackage[normalem]{ulem}       
\usepackage[superscript,biblabel]{cite} 
\usepackage{hyperref}

\hypersetup{
  colorlinks=false,
  linkbordercolor=red,
  citebordercolor=green,
  urlbordercolor=cyan,
  pdftitle={Advances in Materials and Processes for JJs 20251115},
  pdfpagemode=UseOutlines,
}

\newcommand{\dashconnect}[2]{%
  \noindent
  \makebox[1.5cm][l]{#1}%
  \makebox[1cm][l]{\leaders\hbox to 0.5em{\hss-\hss}\hfill}%
  #2\par\vspace{0.5em}
}

\newcommand{\dashconnectraised}[2]{%
  \noindent
  \makebox[0pt][l]{%
    \raisebox{1.3ex}[1.3ex][0pt]{%
      \makebox[1.5cm][l]{#1}%
      \makebox[1cm][l]{\leaders\hbox to 0.5em{\hss-\hss}\hfill}%
    }%
  }%
  \hspace{2.5cm}%
  #2\par\vspace{0.5em}
}

\newcommand{\dashconnectMWraised}[2]{%
  \noindent
  \makebox[1.5cm][l]{#1}%
  \makebox[0pt][l]{%
    \raisebox{1.3ex}[1.3ex][0pt]{%
      \makebox[1cm][l]{\leaders\hbox to 0.5em{\hss-\hss}\hfill}%
      #2%
    }%
  }%
  \par\vspace{0.5em}
}

\usepackage{fancyhdr}
\fancypagestyle{plain}{%
  \fancyhf{}
  \fancyhead[L]{\footnotesize \nouppercase{\rightmark}}
  \fancyhead[R]{\footnotesize WILEY\textendash VCH}
  
  \fancyfoot[C]{\thepage}
}

\begin{document}



\title{Advances in Josephson Junction Materials and Processes \\ Toward Practical Quantum Computing}
\maketitle

\author{Hyunseong Kim,†}
\author{Gyunghyun Jang,†}
\author{Seungwon Jin,†}
\author{Dongbin Shin,}
\author{Hyeon-Jin Shin,}
\author{Jie Luo,}
\author{Akel Hashim,}
\author{Irfan Siddiqi,}
\author{Yosep Kim,*}
\author{Long B. Nguyen,*}
\author{Hoon Hahn Yoon*}

\dedication{$\dagger$ These authors contributed equally to this work.}


\begin{affiliations}
H. Kim, A. Hashim, I. Siddiqi, L. B. Nguyen\\
Applied Mathematics and Computational Research Division, Lawrence Berkeley National Laboratory, Berkeley, CA 94720, USA\\
Department of Physics, University of California, Berkeley, CA 94720, USA\\
Email Address: \url{nbaolong89@gmail.com}


\bigskip
G. Jang,  H.-J. Shin, H. H. Yoon\\
Department of Semiconductor Engineering, Gwangju Institute of Science and Technology (GIST), Gwangju, 61005, Republic of Korea\\
Email Address: \url{hoonhahnyoon@gist.ac.kr}

\bigskip
S. Jin, Y. Kim\\
Department of Physics, Korea University, Seoul, 02841, Republic of Korea\\
Email Address: \url{yosep9201@gmail.com}

\bigskip
D. Shin\\
Department of Physics and Photon Science, Gwangju Institute of Science and Technology (GIST), Gwangju, 61005, Republic of Korea\\
The Max Planck Institute for the Structure and Dynamics of Matter (MPSD) and the Center for Free Electron Laser Science (CFEL), 22761 Hamburg, Germany

\bigskip
J. Luo\\
Anyon Computing Inc., Emeryville, CA 94608, USA
\end{affiliations}

\keywords{Josephson junctions, quantum computing, emerging materials, nanofabrication}

\begin{abstract}
\input{sections/00_abstract}
\end{abstract}

\justifying

\input{sections/10_intro}
\input{sections/20_jj}
\input{sections/30_qcomputing}
\input{sections/31_yield}
\input{sections/32_loss}
\input{sections/33_tunability}
\input{sections/34_footprint}
\input{sections/35_d-wave}
\input{sections/60_fabrication}
\input{sections/61_evap}
\input{sections/62_300mm}
\input{sections/63_epitaxial}
\input{sections/70_outlook_updated}
\input{sections/80_acknowledgements}
\input{sections/81_abbre}


\bibliographystyle{MSP.bst}
\bibliography{references/references}
\input{sections/82_Biography}



\begin{figure}
\textbf{Table of Contents}\\
\medskip
 \includegraphics{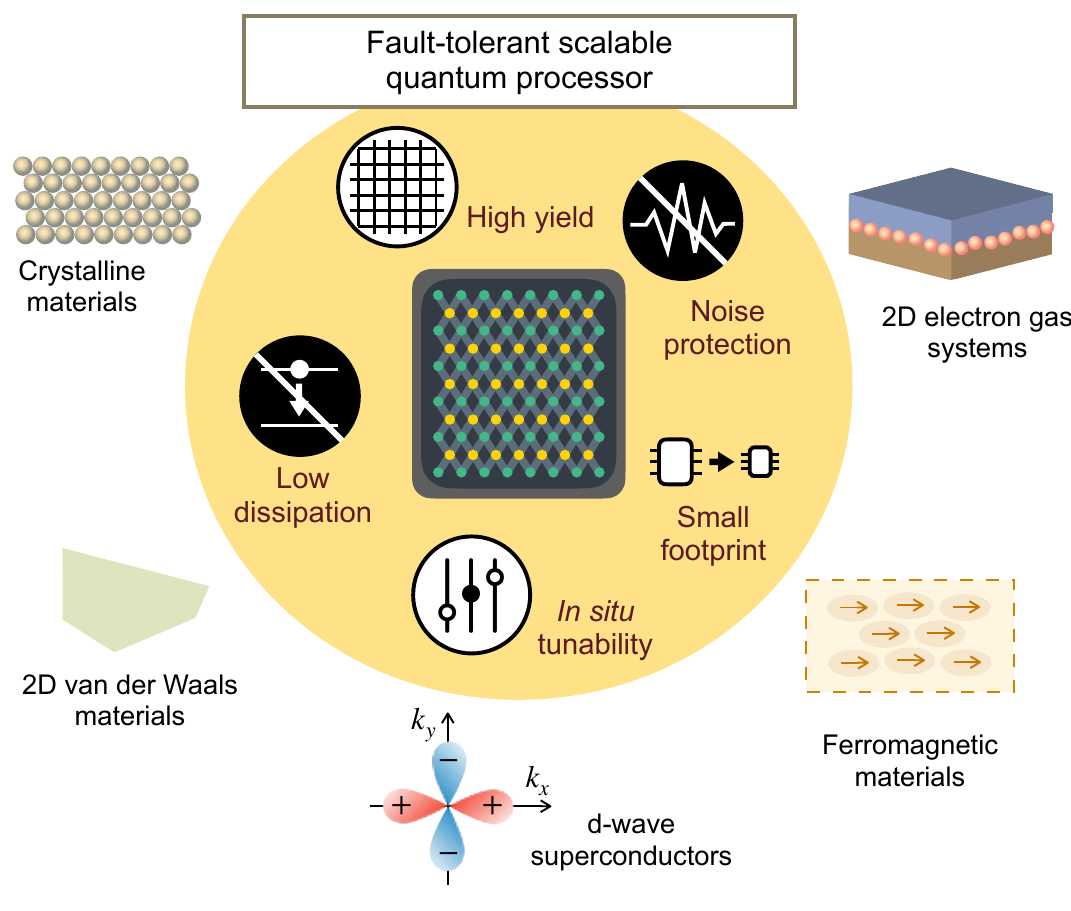}
 \medskip
 \caption*{{\textbf{Table of Contents} Emerging materials for quantum computing, including crystalline materials, two-dimensional van der Waals (2D vdW) materials, two-dimensional electron gas (2DEG) systems, ferromagnets, and unconventional (e.g., d-wave) superconductors, provide new opportunities for engineering next-generation Josephson junctions. By leveraging intrinsic material properties and interfacial design, these platforms enable low dissipation, noise protection, compact device footprint, and in situ tunability, paving the way toward fault-tolerant and scalable superconducting quantum processors.}}
\end{figure}


\end{document}

%% file: sections/00_abstract.tex
\begin{abstract}
The Josephson junction is the fundamental nonlinear building block of superconducting quantum technologies. Its macroscopic quantum tunneling physics underpins superconducting quantum computing, sensing, and communication, but scaling these platforms to utility-scale architectures places increasingly stringent demands on junction materials, interfaces, and fabrication. In quantum computing, these demands include high reproducibility, low dissipation, tunability, compact device footprint, and resilience to noise and defects. This review surveys how advances in materials science, device characterization, and nanofabrication are addressing these challenges and redefining the figures of merit for next-generation Josephson junctions. We also examine the evolution of fabrication strategies, from conventional multi-angle evaporation to foundry-compatible superconducting processes and the integration of emerging junction materials. Progress along these directions will determine how rapidly Josephson junctions move from laboratory-scale components to the foundation of industrial-scale quantum processors.
\end{abstract}

%% file: sections/10_intro.tex
\section{Introduction}

\noindent From the outset of the Industrial Revolution, numerous technological breakthroughs have emerged through cross-pollination among diverse scientific disciplines. 
A paradigmatic example is magnetic resonance imaging, which emerged from the convergence of nuclear physics, chemistry, medicine, mathematics, and computer science.
In the modern era, the development of quantum technologies has also emerged from similar interrelationships among quantum physics, computer science, electrical engineering, and many other fields.
Notably, the convergence of low-temperature electronics, quantum optics, and materials science has given rise to circuit quantum electrodynamics~\cite{blais2021circuit}, which governs the behavior of modern superconducting circuits. Horizontal progress across multiple fronts has propelled superconducting circuits as a leading solid-state platform for scalable quantum technologies. These devices utilize the Josephson junction (JJ), which is the foundational building block of large-scale quantum information processors~\cite{bravyi2022future,mohseni2024build}, the cornerstones of quantum sensing and metrological science~\cite{granata2016nano}, and the core units for modular quantum networks. Moreover, devices based on JJs have led to the development of novel cryogenic microwave components, such as Josephson diodes~\cite{nadeem2023superconducting}, parametric amplifiers~\cite{aumentado2020review}, and integrated circulators~\cite{sliwa2016reconfigurable,navarathna2023passive,fedorov2024nonreciprocity}.

\begin{figure*} 
    \centering
    \includegraphics[width=17.8cm]{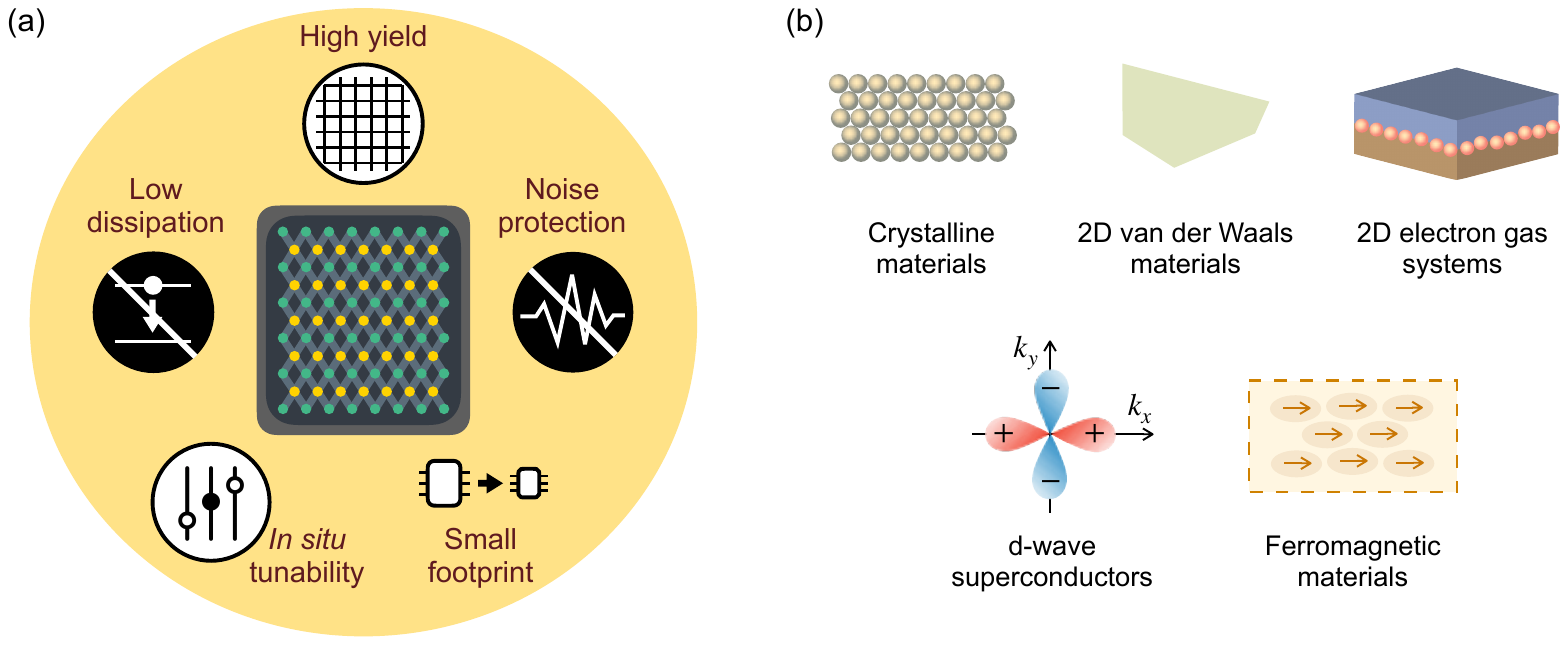}
    \caption{{\textbf{Promising material-oriented strategies for practical superconducting quantum computing.} (a) Current superconducting quantum processors suffer various issues due to fabrication imperfections or system limitations, which can be resolved if qubits become low-dissipative~\cite{bland2025millisecond}, noise-protected~\cite{nguyen2019high}, and frequency tunable~\cite{Qi2018Controlled}. Furthermore, a higher yield of individual qubits on a single processor~\cite{morvan2022optimizing} and a smaller physical size of qubits are highly beneficial to achieve a practical quantum computing era~\cite{wang2022hexagonal}. (b) The performance of superconducting qubits can be enhanced primarily by fabricating highly crystalline metals. Also, opportunities are seen in 2D vdW materials~\cite{balgley2025crystalline}, 2DEG system~\cite{casparis2018superconducting}, d-wave superconductors~\cite{patel2024d}, and ferromagnetic materials~\cite{idzuchi2021unconventional}.}} 
    \label{fig:materials} 
\end{figure*} 

Since the first observation of the Josephson effect using ultra-thin SnO$_\mathrm{x}$ barriers between Sn and Pb superconductors (SCs)~\cite{anderson1963probable}, JJs have been implemented in a wide range of architectures, each instrumental in advancing transformative quantum technologies. Today, the exploration of novel junction materials continues to expand the landscape of quantum hardware, as illustrated in Fig.~\ref{fig:materials}, while innovations in device design and characterization uncover complex phenomena observed in experiments. Concurrently, efforts to scale noisy intermediate-scale quantum (NISQ) devices~\cite{preskill2018quantum} have yielded valuable insights into the challenges and opportunities for building larger systems. The integration of advanced quantum characterization, validation, and verification tools with next-generation Josephson circuits is poised to drive further innovation at the frontiers of quantum technology~\cite{mcrae2021reproducible,hashim2024practical}. For example, systematic monitoring of devices across varied geometries and carefully controlled environments has been key to identifying and mitigating material-related energy losses in superconducting qubits~\cite{de2021materials,mohseni2024build}, enabling continuous improvements in scalability and performance.

In this review, we examine ongoing efforts to advance JJs for practical quantum computing, with a focus on the underlying physical principles and the key performance and integration metrics, including reproducibility, robustness against noise and defects, in-situ tunability, and device footprint. To address these figures of merit, we highlight opportunities enabled by novel junction materials and emphasize the importance of integrating advanced characterization techniques with innovative device architectures supported by creative fabrication methods. We also discuss the nanofabrication challenges inherent in developing next-generation JJ-based devices and the unique opportunities arising from overcoming these hurdles. Although the primary focus is quantum computing, the approaches described here are broadly applicable to other areas of quantum science and technology, including quantum sensing and quantum communication.

This review is structured as follows. First, Section~\ref{section:jj} briefly revisits the Josephson effect, covering the fundamentals needed to understand the nonlinear behaviors of JJs. Section~\ref{section:computing} surveys the key figures of merit crucial to the construction of superconducting quantum processors, encompassing tunnel barrier uniformity (\ref{section:yield}), energy loss mechanisms associated with the junctions (\ref{section:tls}), enhanced tunability (\ref{section:tunability}), reduced device footprint (\ref{section:footprint_scalability}), and cutting-edge ideas about robust qubits derived from d-wave SCs or JJs with a ferromagnetic insulating interlayer (\ref{section:d-wave}). Section~\ref{section:fab} focuses on advanced nanofabrication techniques, highlighting large-scale integration and process compatibility. Finally, Section~\ref{section:summary} presents an outlook on the future of JJs in quantum processors, discussing emerging material platforms, scalable fabrication strategies, and insights from the historical evolution of the CMOS (complementary metal-oxide-semiconductor) integrated circuit (IC) chip industry.


%% file: sections/20_jj.tex
\section{Revisiting the Josephson Effect}
\label{section:jj}

\begin{figure*} 
    \centering
    \includegraphics[width=17.8cm]{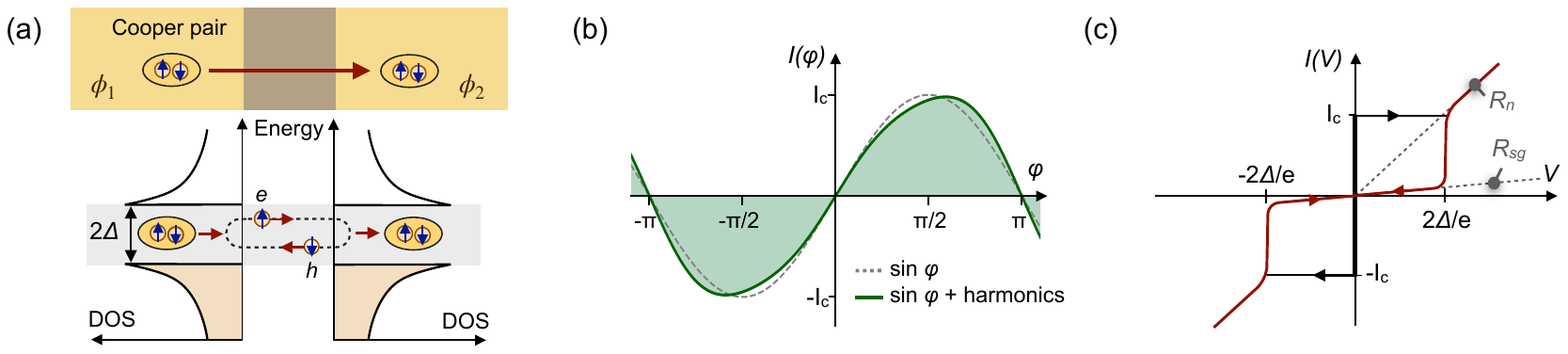}
    \caption{{\textbf{Transport mechanisms, CPR, and I–V characteristics in JJs.} (a) Charge transport mechanisms in superconductor–insulator–superconductor (SIS) and superconductor–normal metal–superconductor (SNS) junctions. In SIS junctions, Cooper pairs tunnel coherently through the insulating barrier. In SNS junctions, phase-coherent transport occurs via Andreev reflection, leading to the formation of ABS in the normal region, which carries the supercurrent~\cite{Tinkham1996Introduction}. (b) CPR of JJs. SIS junctions in the tunneling limit exhibit a sinusoidal CPR, $I(\varphi) = I_c \sin\varphi$. In contrast, SNS junctions with high transparency support Andreev bound states whose nonlinear phase dispersion leads to nonsinusoidal CPR containing higher harmonics~\cite{beenakker1991universal}. (c) Nonlinear I-V characteristics of JJs. The normal-state resistance $R_n$ is defined from the slope at voltages above the superconducting gap, whereas the subgap resistance $R_{\mathrm{sg}}$ characterizes dissipative transport within the superconducting gap region~\cite{devoret2013superconducting}.}} 
    \label{fig:Josephson} 
\end{figure*}

\noindent In 1962, Brian Josephson predicted the dissipationless flow of a supercurrent between two superconducting electrodes separated by a thin insulating barrier, a phenomenon now known as the \emph{Josephson effect} [(Fig.~\ref{fig:Josephson}(a)]~\cite{Josephson1962}. At the time, quantum tunneling was well established for normal (non-superconducting) electrons; Josephson’s key insight was that Cooper pairs could likewise tunnel coherently across a weak link. The subsequent experimental verification of this prediction~\cite{anderson1963probable,josephson1974discovery} led to Josephson being awarded the 1973 Nobel Prize in Physics. A particularly striking aspect of the Josephson effect is that the supercurrent depends on the relative phase difference between the superconducting order parameters, a concept that was both unexpected and conceptually profound at the time. In modern times, engineering the current-phase relationship in Eq.~\eqref{eq:JJ_CPR}, its functional form, symmetry, and nonlinearity has become a central focus in developing novel JJs, driving fundamental research and technological advancements.

The Josephson effect arises from the macroscopic quantum coherence of the superconducting condensate. Each SC is characterized by a macroscopic wavefunction $\psi = \sqrt{n}e^{i\phi}$, where $n$ is the Cooper-pair density and $\phi$ is the phase. The gradient of phase governs the supercurrent flow, described by the probability current density $\Vec{J} \propto Im[\psi^*\nabla\psi]$. For a uniform Cooper-pair density, this simplifies to $\Vec{J} \propto \nabla\phi$, showing that the supercurrent is proportional to the phase gradient. Therefore, when two SCs are weakly coupled through a tunneling barrier, the phase difference $\delta\phi$ drives a dissipationless supercurrent. Due to the single-valuedness of the condensate wave function, this current is a periodic function of $\delta\phi$ with a period equal to $2\pi$. Furthermore, time-reversal symmetry (TRS) ensures that the current is an odd function of $\delta\phi$, which results in the characteristic sinusoidal current-phase relation (CPR),
\begin{equation}\label{eq:JJ_CPR}
    I = I_c\sin(\varphi),
\end{equation}
where $I_c$ is the critical current and $\varphi\equiv \delta \phi$. 

In addition to the CPR, the Josephson effect also encompasses a dynamical component that links the time derivative of the phase difference to the voltage across the junction. This is commonly expressed as the second Josephson relation,
\begin{equation}\label{eq:JJ_phase_voltage}
    \frac{d\varphi}{dt} = \frac{2\pi}{\Phi_0}V,
\end{equation}
which follows directly from the Schrödinger equation and Faraday's law of electromagnetic induction. The dynamic aspect of the Josephson effect establishes a direct connection between the quantum phase and observable electrical quantities, serving as the foundation for applications such as voltage standards. Furthermore, Eq.~\eqref{eq:JJ_phase_voltage} directly leads to the Josephson energy-phase relation,
\begin{equation}\label{eq:JJ_energyphase}
\begin{split}
    E(\varphi) &= \int I(\varphi)V dt = \int I_c\sin(\varphi) \frac{\Phi_0}{2\pi}d\varphi \\
    &= -\frac{\Phi_0 I_c}{2\pi}\cos\varphi = -E_\mathrm{J}\cos\varphi,
\end{split}
\end{equation}
where $E_\mathrm{J}$ is the so-called \emph{Josephson energy}. Eq.~\eqref{eq:JJ_energyphase} is fundamental to understanding and engineering systems in circuit quantum electrodynamics (cQED). It highlights the JJ's coherent and nonlinear properties, which are fundamental to its role in enabling a broad spectrum of quantum technologies.

While Eq.~\eqref{eq:JJ_energyphase} captures the fundamental sinusoidal dependence of the Josephson energy on the superconducting phase difference, it reflects an idealized tunneling limit in which only the first harmonic is significant. In general, in normal materials such as insulators, normal metals, and semiconductors (SMs), the CPR can exhibit substantially higher-order Josephson harmonics, depending on the microscopic properties of the junction. According to microscopic theory, the dissipationless Josephson current arises from the lower level \textit{Andreev bound state} (ABS), which forms due to multiple Andreev reflections at the interfaces between the SCs and the normal barrier as shown in Fig.~\ref{fig:Josephson}(a)~\cite{Andreev1964Thermal, martinis2004superconducting}. In an Andreev reflection, an electron incident on the interface is retroreflected as a hole, effectively transferring a Cooper pair across the interface. In the short junction regime, repeated reflections that constructively interfere within the junction give rise to phase-dependent bound states with energies given by Eq.~\eqref{eq:ABS_Energy}
\begin{equation}\label{eq:ABS_Energy}
    E_\mathrm{ABS}(\varphi) = \pm\Delta \sqrt{1 - \tau \sin^2(\varphi/2)},
\end{equation}
where \(\tau\) is the conduction channel's transmission probability, or its so-called \emph{transparency}, and \(\Delta\) is the superconducting gap~\cite{Beenakker1991Josephson}. In the low temperature limit, the Josephson current is carried by the lower (negative) energy ABS. Summing over all conduction channels and applying the second Josephson relation, we arrive at the generalized CPR,
\begin{equation}\label{eq:I_ABS}
    \begin{split}
         I(\varphi) &= \sum_i \frac{e\Delta}{\hbar} \times \frac{\tau_i \sin(\varphi)}{2\sqrt{1 - \tau_i \sin^2(\varphi/2)}} \\
         &= \sum_i\sum_{k} I_{ki} \sin(k\varphi),
    \end{split}
\end{equation}
where the harmonic decomposition comes from the binomial series expansion (Appendix B). Here, the higher harmonics corresponding to $k>1$ are effective manifestations of multi-Cooper-pair tunneling processes~\cite{willsch2024observation}. In the low transparency limit, \(\tau_i \ll 1\), the CPR reduces to the sinusoidal form of Eq.~\eqref{eq:JJ_CPR}. However, higher-order harmonics become significant for highly transparent channels, indicating a strongly anharmonic CPR [Fig.~\ref{fig:Josephson}(b)]. By tuning transparency through material choice, junction geometry, or electrostatic gating, one can sculpt the CPR to enhance qubit performance (see Section~\ref{section:d-wave})~\cite{Golubov2004current}.

While the CPR is naturally expressed in terms of the transmission probabilities $\tau_i$ of individual conduction channels, these microscopic quantities are not directly accessible experimentally. Instead, the overall transparency of the junction can be inferred from the normal-state resistance $R_n$ through the Landauer formalism. The key properties of a JJ must be accurately characterized before it can be integrated into a quantum application. A fundamental and practical method for this characterization is to extract the critical current $I_c$ from the normal-state resistance $R_n$, measured right after the superconducting gap [Fig.~\ref{fig:Josephson}(c)]. The resistance of a quantum conductor is related to the aggregate transparency of the conduction channels through Landauer’s formula~\cite{Landauer1957}:
\begin{equation}\label{eq:Landauer}
    \frac{1}{R_n} = \sum_i\frac{e^2}{\pi\hbar} \tau_i.
\end{equation}
Applying Eq.~(\ref{eq:Landauer}) to the CPR in Eq.~(\ref{eq:I_ABS}), and considering the low-transparency limit, leads to the well-known Ambegaokar-Baratoff relation~\cite{Ambegaokar1963}:
\begin{equation}
    I_c = \frac{\pi\Delta}{2eR_n}.
\end{equation}
This result directly estimates the JJ's critical current from normal-state resistance measurements, enabling efficient yield characterization of superconducting quantum processors (Subsection~\ref{section:yield}). Importantly, it offers a noninvasive diagnostic tool that reduces the need for extensive cryogenic cycling, streamlining the fabrication and testing of large-scale superconducting circuits. Notably, this relation is exact only in the tunneling limit, with deviations of up to a factor of 2× in the ballistic limit~\cite{willsch2024observation}, which provides a diagnostic of the transmission distribution.

Cryogenic characterization techniques provide essential insight into the transport properties of JJs under their actual operating conditions. One of the most informative tools is the current-voltage (I–V) measurement, which can be used to determine the critical current, the point at which a finite voltage develops across the junction, corresponding to the breakdown of the dissipationless supercurrent~\cite{clarke2004squid}. For tunnel junctions, if the junction exhibits leakage current, i.e., quasiparticle transport below the superconducting gap, a finite subgap voltage is characterized by the subgap resistance shown in Fig.~\ref{fig:Josephson}(c). This resistance is inversely proportional to the population of thermally excited quasiparticles, which act as sources of decoherence in quantum devices~\cite{Catelani2011}. Consequently, a considerable subgap resistance is crucial for quantum applications such as superconducting qubits (Section~\ref{section:computing}).

The detailed shape of the I–V contour offers additional information about the junction’s internal dynamics~\cite{Tinkham1996Introduction, Baumgartner2021, Shapiro1963, Vanneste1988}. For example, hysteresis in the I–V curves typically signals an underdamped junction, where a significant normal-state resistance and junction capacitance contribute to inertial phase dynamics. Asymmetries that indicate a shifted or distorted CPR can be used to introduce nonreciprocal elements. Furthermore, when the junction is subjected to an oscillating voltage of frequency $\omega$, the I–V curve exhibits a series of plateaus known as Shapiro steps, occurring at voltages $V = \nu \hbar \omega/2e$, where $\nu\in \mathbb{Z}$. In junctions exhibiting higher-order Josephson harmonics, fractional Shapiro steps appear, indicating multi-Cooper-pair tunneling. As such, I–V characterization remains a cornerstone technique for probing JJs. Additional techniques, such as measuring the modulation of the critical current in response to an applied magnetic field or embedding a junction within a loop containing another well-characterized junction, provide valuable insights into both the spatial uniformity and the CPR~\cite{Dynes1971,Fominov2022}. With a solid understanding of the JJ’s transport properties, we now focus on its diverse quantum applications, which leverage its dissipationless transport and nonlinear behavior.
\vspace{1em}

%% file: sections/30_qcomputing.tex
\begin{table*}[t]
    \centering
    \caption{Performance metrics for quantum computing and some relevant challenges for improvement.}
    \label{tab:Key_properties}

    \begin{tabular}{ >{\raggedright}p{0.25\textwidth} 
                    >{\raggedright}p{0.25\textwidth} 
                    >{\raggedright\arraybackslash}p{0.45\textwidth} }
    \hline\hline
    \multicolumn{1}{>{\centering\arraybackslash}p{0.25\textwidth}}{\textbf{Figures of Merit}} & 
    \multicolumn{1}{>{\centering\arraybackslash}p{0.25\textwidth}}{\textbf{Challenges}} & 
    \multicolumn{1}{>{\centering\arraybackslash}p{0.45\textwidth}}{\textbf{Opportunities}} \\ 
    \hline\hline
    
    \begin{itemize}[leftmargin=1.2em, label=\textbullet, nosep]
        \item Reproducibility
        \item Yield    
        \item Long-term junction stability
    \end{itemize}
    & 
    \begin{itemize}[leftmargin=1.2em, label=\textbullet, nosep]
        \item Varying junction thickness
        \item Varying junction area
        \item Line-edge roughness
        \item Aging effect
    \end{itemize}
    &
    \begin{itemize}[leftmargin=1.2em, label=\textbullet, nosep]
        \item Advanced fabrication with precise barrier thickness
        \item Novel junction materials
        \item Post-fabrication corrections
        \item Passivation and thermal annealing
    \end{itemize} \\[-1.7ex] 
    \hline
        
    \begin{itemize}[leftmargin=1.2em, label=\textbullet, nosep]
        \item Energy dissipation
    \end{itemize}
    & 
    \begin{itemize}[leftmargin=1.2em, label=\textbullet, nosep]
        \item Two-level systems
        \item Quasiparticle tunneling
    \end{itemize}
    & 
    \begin{itemize}[leftmargin=1.2em, label=\textbullet, nosep]
        \item Low-loss crystalline tunnel barriers
        \item 2D vdW materials and their heterostructures
        \item Gap-engineered electrodes

    \end{itemize} \\[-1.7ex] 
    \hline
    
    \begin{itemize}[leftmargin=1.2em, label=\textbullet, nosep]
        \item \textit{In situ} tunability
    \end{itemize}
    & 
    \begin{itemize}[leftmargin=1.2em, label=\textbullet, nosep]
        \item Heat load and crosstalk due to flux tuning
        \item Footprint of flux loops
    \end{itemize}
    & 
    \begin{itemize}[leftmargin=1.2em, label=\textbullet, nosep]
        \item Gate-tunable 2DEG systems
        \item Gate-tunable 2D vdW JJs
    \end{itemize} \\[-1.7ex] 
    \hline
    
    \begin{itemize}[leftmargin=1.2em, label=\textbullet, nosep]
        \item Device footprint
    \end{itemize}
    & 
    \begin{itemize}[leftmargin=1.2em, label=\textbullet, nosep]
        \item Large-area circuit elements
    \end{itemize}
    &
    \begin{itemize}[leftmargin=1.2em, label=\textbullet, nosep]
        \item Merge-element transmon qubits 
        \item High-$\kappa$ 2D vdW tunnel barriers
    \end{itemize} \\[-1.7ex] 
    \hline
    
    \begin{itemize}[leftmargin=1.2em, label=\textbullet, nosep]
        \item Noise-protected encoding
    \end{itemize}
    & 
    \begin{itemize}[leftmargin=1.2em, label=\textbullet, nosep]
        \item Tuning Cooper-pair parity
    \end{itemize}
    &
    \begin{itemize}[leftmargin=1.2em, label=\textbullet, nosep]
        \item Twisted d-wave flakes
        \item Stacking d-wave and s-wave flakes
        \item $\pi$-junctions with ferromagnet interlayers
    \end{itemize} \\[-1.7ex] 
    \hline\hline
    \end{tabular}
\end{table*}

\section{Quantum Computing}
\label{section:computing}

\noindent Superconducting circuits are among the most promising platforms for realizing practical quantum computers. Superconducting qubits leverage the well-established principles of superconductivity and the Josephson effect to create quantum systems that can be controlled in a scalable manner~\cite{devoret2013superconducting}. Remarkable progress is evident through the growing qubit counts and increasingly complex quantum operations, which have enabled milestones such as demonstrating quantum supremacy and quantum error correction~\cite{Arute2019supremacy, acharya2025}. There is considerable effort to achieve a meaningful computational speedup over classical computing for practical applications, a critical benchmark often referred to as \emph{quantum utility}~\cite{kim2023evidence}. In parallel with advances in quantum computing hardware, significant efforts are directed toward developing practical quantum algorithms suited for current and near-term devices~\cite{martyn2021,katabarwa2024early,dalzell2025quantum} and quantum simulations~\cite{altman2021quantum}, with a particular emphasis on scientific applications~\cite{alexeev2021quantum,bauer2023quantum}.

\begin{figure*}[t]
    \centering
    \includegraphics[width=17.8cm]{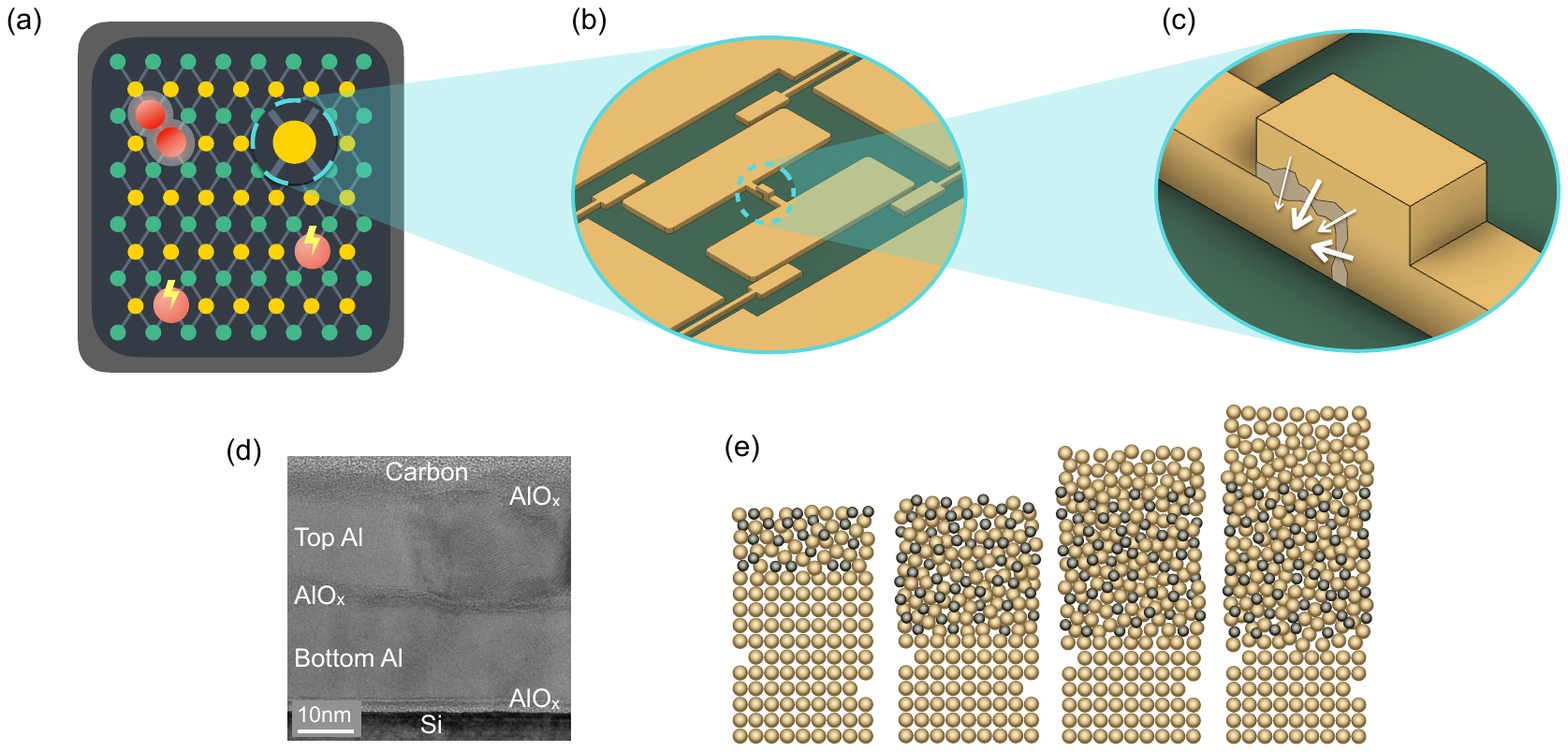}
    \caption{\textbf{JJ imperfections.} (a) Schematic of a superconducting quantum processor, highlighting defective outlier qubits caused by frequency collisions and defects due to interaction with the lossy environment. (b) A JJ connecting two large superconducting electrodes, forming a transmon qubit. Smaller structures adjacent to the pads enable capacitive coupling to external circuitry. These sizes are not to scale. (c) Illustration of spatial variations in supercurrent density (white arrows) arising from nonuniform junction thickness and surface roughness along the tunneling barrier~\cite{zeng2015direct,zeng2016atomic}. (d) Transmission electron microscopy (TEM) image of an Al/Al$\mathrm{O_x}$/Al junction fabricated on a silicon substrate. The few-nanometers Al$\mathrm{O_x}$ layer with varying thickness largely determines the Josephson energy $E_\mathrm{J}$~\cite{kim2022effects}. (e) Model of atomic structure evolution during Al/Al$\mathrm{O_x}$/Al junction formation, showing the oxidation and deposition process on a crystalline aluminum bottom electrode (yellow: aluminum, black: oxygen). Molecular dynamics simulation with the Streitz–Mintmire potential enables analysis of structural properties during junction formation under varying conditions such as temperature, bond angle, and oxygen pressure~\cite{cyster2021simulating}.}
    \label{fig:Imperfection}
\end{figure*}

The core of superconducting qubit technology is currently the Al/AlO$_\mathrm{x}$/Al tunnel junction. These junctions are the building blocks of contemporary qubits, providing nonlinear inductance that enables the manipulation of quantum states. Combining several junctions with a toolbox of superconducting capacitors and inductors enables the construction of various qubits with different properties~\cite{devoret2013superconducting,siddiqi2021engineering}. The transmon, as shown in Fig.~\ref{fig:Imperfection}(b), is the most common type of superconducting qubit due to its superior coherence properties and ease of fabrication. The qubit features a JJ shunted by a large capacitor, functioning as a nonlinear oscillator. Adding a second JJ in parallel creates a nonlinear oscillator that can tune the magnetic flux threading the loop of junctions, known as a \emph{superconducting quantum interference device} (SQUID).

The success of superconducting transmon qubits and qudits~\cite{blok2021quantum,goss2022high,liu2023performing,fischer2023universal,nguyen2024empowering} has been driven by sustained efforts to improve coherence through horizontal approaches, such as reinforced shielding and filtering~\cite{glazman2021bogoliubov,diamond2022distinguishing,connolly2024coexistence}, material purification through advanced fabrication techniques~\cite{de2021materials}, packaging~\cite{huang2021microwave}, and optimization of qubit geometry~\cite{paik2011observation,wang2015surface,siddiqi2021engineering,pan2022engineering}. Interestingly, the transmon's simplicity makes it an ideal vehicle for probing the surrounding noisy environment and examining general decoherence mechanisms. Once perceived as an insurmountable obstacle, the strong coupling of superconducting qubits to their imperfect bath has proven valuable in exploring microscopic material defects that may affect other solid-state platforms~\cite{muller2019towards}. These newfound insights into noise properties have deepened our understanding of decoherence and play a pivotal role in developing new circuit platforms designed for superior noise resilience~\cite{gyenis2021experimental,nguyen2022blueprint}.

As the number of qubits in quantum processors increases, solutions that work effectively at small scales may no longer suffice for large-scale systems. Achieving high-performance quantum computation at scale will require fundamentally new approaches. One critical challenge is that the device's worst-performing qubit often constrains a quantum processor's performance~\cite{mohseni2024build,acharya2025}. While significant progress has been made in improving average qubit performance, prioritizing uniformity across the entire device is now essential to ensure consistent reliability. Another pressing need is the development of qubits with smaller footprints, as chip size imposes inherent physical limitations. Reducing the qubit size will enable higher qubit densities without compromising functionality. Moreover, implementing additional protection mechanisms could substantially improve qubit coherence times, further advancing device performance. In the following discussion, we explore how innovative engineering methodologies for JJs could address these challenges and meet the demands of next-generation superconducting quantum processors.

%% file: sections/31_yield.tex
\subsection{Yield \& Reproducibility} 
\label{section:yield}

\noindent Superconducting quantum processors, shown in Fig.~\ref{fig:Imperfection}(a), require the constituent qubits to be spectrally distributed to suppress spurious interactions and enable individual addressability~\cite{hertzberg2021laser}. This underscores the necessity of scaling with exceptionally high JJ fabrication precision. While device yield may vary across different platforms~\cite{nguyen2022blueprint}, we explore here the transmon qubit as a representative example, as depicted in Fig.~\ref{fig:Imperfection}(b). Its computational transition frequency is given by
\begin{equation}
\omega_{01}/2\pi \approx \sqrt{8 E_\mathrm{J} E_\mathrm{C}} - E_\mathrm{C},
\end{equation}
where $E_\mathrm{C} = e^2/2C$ is the charging energy set by the shunt capacitance $C$, and $E_\mathrm{J} = \Phi_0 I_c/2\pi$ is the Josephson energy determined by the critical current of the junction (see Eq.~(\ref{eq:JJ_energyphase})). While capacitive elements with relatively large dimensions can be fabricated with high reproducibility, JJs are highly susceptible to fabrication-induced variations. For instance, a 3\% fluctuation in $I_c$ results in an approximately 1.5\% change in $\omega_{01}$, corresponding to a 75~MHz shift for a transmon qubit designed to operate at 5~GHz, which is substantial relative to typical qubit frequency spacings of tens to a hundred megahertz~\cite{koch2007charge}. 

Thus, achieving high precision in JJ fabrication is essential for maintaining device yield in large-scale superconducting quantum processors~\cite{morvan2022optimizing}.
Although frequency-tunable superconducting qubits can mitigate frequency collisions, precise control over $I_c$ remains essential for maintaining optimal qubit performance at noise-resilient operating points~\cite{nguyen2022blueprint}.
Moreover, while post-fabrication frequency tuning strategies can compensate for some fabrication-induced variability~\cite{kim2022effects,zhang2022high,odeh2025non}, improving the reproducibility of JJs remains indispensable for achieving scalable quantum processors.

The junction's critical current $I_c$ is primarily determined by its dimensions and the material structure of the tunnel barrier. Advances in shadow evaporation techniques, combined with optimized oxidation processes and bias corrections, have reduced variations in $I_c$ to within $\sim2\%$ across wafers~\cite{kreikebaum2020improving,moskalev2023optimization,pishchimova2023improving,osman2023mitigation,muthusubramanian2024wafer}.
However, conventional multi-angle evaporation faces inherent challenges, including the mechanical instability of resist bridges, increased line-edge roughness, and radial nonuniformities introduced by angled deposition, which may fundamentally limit achievable junction reproducibility. To overcome these limitations, fabrication approaches compatible with optical lithography and orthogonal dry etching have been developed~\cite{tolpygo2014fabrication,verjauw2022path,stehli2020coherent,anferov2024improved,ke2025scaffold}, providing sub-100-nm critical dimension control and enabling scalable fabrication on 300-mm wafers. Such etch-based processes have achieved uniformity in $I_c$ comparable to that of advanced shadow evaporation techniques~\cite{van2024advanced}.
A more detailed description of these fabrication techniques and their limitations is provided in Section~\ref{section:fab}. 

Despite these advances, downstream effects inherent to junction physics, secondary consequences of the underlying mechanisms, such as frequency shifts and parameter drifts, continue to limit the precision of frequency targeting.
Even with tight control over the junction area and the thickness of the tunnel barrier, microscopic nonuniformities~\cite{zeng2015direct,zeng2016atomic}, such as variations in local thickness in the junction or roughness in the electrode and barrier layers [Fig.~\ref{fig:Imperfection}(c,d)], can induce substantial fluctuations in local critical current density, ultimately impacting junction consistency.
Careful management of these microscopic effects, such as narrowing the distribution of transmission distributions, is therefore crucial for maintaining fabrication quality.
Further improvements through oxidation optimization~\cite{fritz2019structural,chen2023optimization} and epitaxial growth~\cite{qiu2020fabrication,nakamura2011superconducting,kim2021enhanced} allow for atomic-scale control of tunnel barrier properties.
As illustrated in Fig.~\ref{fig:Imperfection}(e), fabricating highly crystalline bottom electrodes facilitates uniform oxidation and top electrode deposition~\cite{oh2006elimination,cyster2021simulating}, thereby improving junction uniformity.
Furthermore, post-fabrication treatments, such as thermal annealing~\cite{pavolotsky2011aging,koppinen2007complete} and surface oxide passivation~\cite{bilmes2021situ,bal2024systematic} have been shown to mitigate frequency drift and enhance the long-term stability of the device.

Beyond the frequency targeting achieved through process control, the intrinsic material quality of JJs plays a critical role in determining the scalability and yield of superconducting quantum processors.
Localized defects near the junction, including two-level systems (TLS) in tunnel barriers, quasiparticle generation from external radiation, and dielectric loss at material interfaces, introduce variability in qubit relaxation times and coherence properties (see Section \ref{section:tls}). Tightening the temporal variations in fidelity is crucial for realizing fault-tolerant quantum computation, as even a small number of low-fidelity qubits can compromise the performance of quantum error correction codes~\cite{mohseni2024build,hashim2024practical}. Addressing this challenge requires integrating efficient quality control strategies into fabrication processes. In particular, early detection of fabrication-induced defects is essential for maintaining high device yield across large wafers. To this end, incorporating \textit{in situ} optical defect inspection, a well-established method in CMOS process flow, provides a promising route to identify and mitigate fabrication-induced defects early in the process~\cite{zhu2022optical}. In conjunction with cryogenic measurement feedback, new junction processes may also benefit from high-resolution material imaging techniques in Fig.~\ref{fig:Characterization}, such as transmission electron microscopy, scanning electron microscopy, and atomic force microscopy when deployed at scale.

%% file: sections/32_loss.tex
\subsection{Energy Dissipation at the Junctions}
\label{section:tls}

\subsubsection{Two-Level Systems in Tunnel Barriers}

\begin{figure*}
    \centering
    \includegraphics[width=17.8cm]{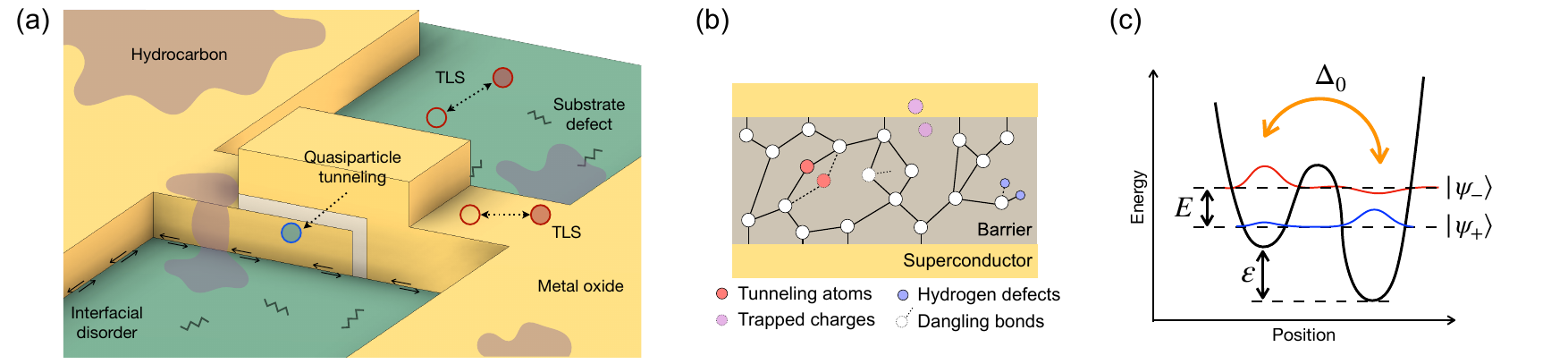}
    \caption{\textbf{Energy loss sources around the junction and double well potential model of TLS.} (a) Schematic illustration of dominant loss mechanisms surrounding a transmon qubit. Material imperfections, such as hydrocarbon residues, interfacial disorder, amorphous metal–oxide layers, and substrate defects, host two-level systems (TLS) that interact with the electric field within the qubits and give rise to dielectric loss. Additional losses arise from quasiparticle tunneling across the junction~\cite{grabovskij2012strain}. (b) Microscopic depiction of two-level systems within the tunnel barriers between the superconducting electrodes~\cite{lisenfeld2019electric,lisenfeld2023enhancing}. Fluctuations in their configurations induce qubit-frequency noise, spectral diffusion, and relaxation hot spots. (c) Double-well potential representation of a two-level system (TLS) in the standard tunneling model. An atom or atomic configuration occupies two nearly equivalent minima separated by a barrier, allowing quantum tunneling between localized states at low temperatures. This tunneling forms hybridized eigenstates $|\psi_{+}\rangle$ and $|\psi_{-}\rangle$ with a finite energy splitting determined by the asymmetry between the wells and the tunneling strength. Such TLS defects can resonantly interact with a qubit and contribute to dielectric loss~\cite{lubchenko2007microscopic}.}
    \label{fig:TLS}
\end{figure*}

\noindent Solid-state defects, commonly referred to as ``two-level systems'' (or TLSs), have garnered attention as a critical factor affecting the performance of superconducting qubits. Their behavior at low temperatures has long been a fascinating topic of research in condensed matter physics~\cite{muller2019towards}. Although a universal consensus on their microscopic origins remains elusive, the observed properties of glassy materials are often used to attribute the existence of TLS to the absence of long-range atomic order in the amorphous state of these materials, as shown in Fig.~\ref{fig:TLS}(a,b).


Surprisingly, TLSs display remarkable similarities across different chemical compositions and material arrangements~\cite{dutta1981low,paladino2014,muller2019towards}. These properties are generally explained by the standard tunneling model, described in Fig.~\ref{fig:TLS}(c)~\cite{lubchenko2007microscopic}. In this model, the two energetically similar states of a TLS are separated by a finite potential barrier, forming a double-well potential, whose Hamiltonian can be written as $H_{TLS}=\frac{1}{2}\varepsilon\sigma_z+\frac{1}{2}\Delta_0\sigma_x$, with an eigenstate splitting $E_\mathrm{TLS}=\sqrt{\varepsilon^2+\Delta_0^2}$. At sufficiently low temperatures, thermal activation between the two localized states is suppressed, and quantum tunneling hybridizes them into eigenstates split by $E_\mathrm{TLS}$. These defects can then exchange energy with electric fields, phonons, other TLSs within the surrounding materials, or the host qubit, thereby contributing to dielectric loss, dephasing, and spectral diffusion~\cite{muller2015interacting,faoro2015interacting,lisenfeld2015observation,chen2024phonon,odeh2025non}.

TLSs exhibit a distinct signature that provides valuable insights into their behavior~\cite{muller2019towards}. For instance, their spectral signatures become thermally broadened or less distinct at elevated temperatures and energies~\cite{crowley2023disentangling} and are tunable with external factors like electric fields~\cite{lisenfeld2019electric,lisenfeld2023enhancing,chen2025scalable} and strain~\cite{grabovskij2012strain}. In the dressed frame of a superconducting qubit, TLSs can couple to the qubit either linearly or nonlinearly~\cite{abdurakhimov2022identification}, offering a unique opportunity to probe their spectral distribution and distinguish TLS-induced loss from other dissipation sources~\cite{de2020two,connolly2024coexistence}. These signatures enable quantitative studies of how TLSs limit qubit coherence and provide experimental handles for identifying and mitigating defect-induced loss.

TLSs are pervasive in solid-state devices, appearing in regions such as the bulk substrate, amorphous surfaces and interfaces, and within tunnel junction barriers~\cite{wang2015surface,ganjam2024surpassing}. By transforming the transmon into a spectral TLS sensor~\cite{klimov2018fluctuations}, recent experiments have combined strain tuning and electric fields as control knobs to pinpoint the locations of TLS and their coupling strengths to the host qubit~\cite{lisenfeld2019electric, bilmes2021situ, bilmes2022probing}. In addition, varying the spatial sensitivity to TLSs in multimode planar resonators and transmons has enabled researchers to separate the contributions of these components across various parts of the device~\cite{lei2023characterization, ganjam2024surpassing}. Measurements at different temperatures~\cite{crowley2023disentangling} and driven frames~\cite{abdurakhimov2022identification,nguyen2024programmable} can also distinguish normal TLSs from other effects, such as quasiparticle loss~\cite{de2020two}.

Research and development (R\&D) in JJ materials and superconducting qubit circuits has primarily focused on reducing TLS levels in amorphous bulk dielectric layers. For example, introducing Ta- or annealed-sapphire substrates has significantly mitigated bulk dielectric losses. Notable advances include the identification of surface oxides that form in various superconducting materials and the correlation of these findings with different substrate purification processes~\cite{quintana2014characterization,wang2015surface,place2021new,premkumar2021microscopic,altoe2022localization,crowley2023disentangling,ganjam2024surpassing,bland2025millisecond}.





In addition to electric-dipole coupling, some defects in bulk crystalline dielectrics may couple to qubits through strain- or piezoelectric-mediated channels~\cite{zhang2024acceptor}. TLSs can also form from adsorbates, leading to pure dephasing in flux-tunable devices. Moreover, TLSs that reside at the junction interface are strongly coupled to the host qubit due to the geometry of the transmon, which concentrates the electric field across the junction~\cite{simmonds2004decoherence,neeley2008process,ganjam2024surpassing}. Therefore, fluctuating TLS residing within the tunneling barrier and surrounding dielectric can cause substantial variations in qubit performance~\cite{ganjam2024surpassing}, contributing to device-to-device variability and temporal instability in qubit performance~\cite{mohseni2024build}.

As several other decoherence channels have been progressively reduced in leading transmon platforms, TLS-resilient JJ materials have become increasingly important as quantum processors scale up. Compared to amorphous materials, the dielectric loss ($\tan\delta$) of crystalline materials like sapphire~\cite{ganjam2024surpassing,zhang2024acceptor} has inspired efforts to develop crystalline tunnel barriers. One promising approach is to grow a thin epitaxial insulating layer, such as Al$_2$O$_3$~\cite{oh2006elimination,kline2011sub,fritz2019optimization,wu2020vacuo} or nitride-based insulators~\cite{nakamura2011superconducting,weides2011coherence,wang2013high,qiu2020fabrication,kim2021enhanced,bhatia2024enabling}, between superconducting electrodes made from compatible materials. Preliminary TLS spectroscopy experiments have shown a reduced density of TLS in junctions~\cite{oh2006elimination,hung2022probing}.

\begin{figure*}
    \centering
    \includegraphics[width=17.8cm]{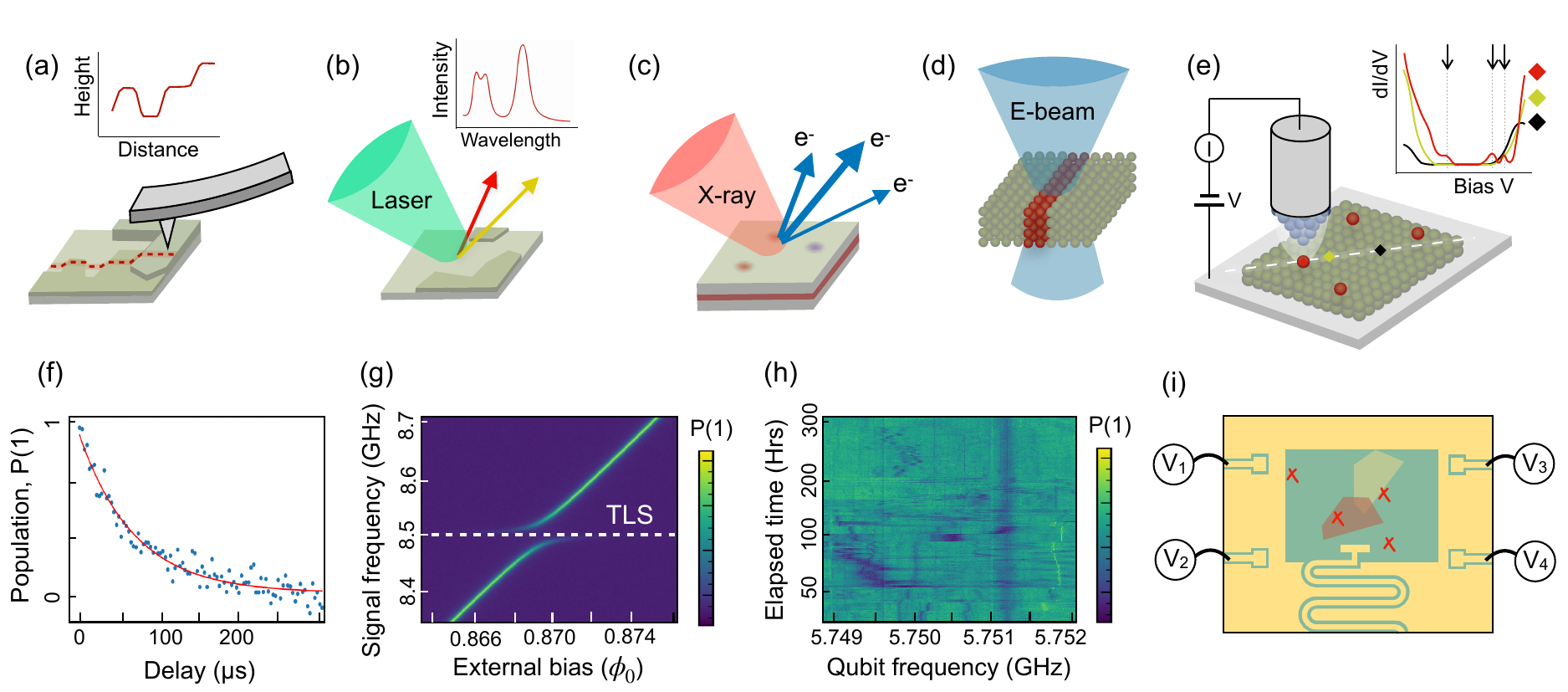}
    \caption{\textbf{Material characterization for low-loss JJs and TLS evaluation of superconducting qubits.} (a) AFM for probing surface morphology, roughness, and thickness uniformity. (b) Raman and PL spectroscopies for probing layer number, strain, doping, phase purity, and transfer-induced disorder. Red and yellow arrows represent scattered and emitted light, respectively. (c) XPS for analyzing elemental composition, oxidation states, and stoichiometry. (d) STEM for probing crystallinity, atomic structure, and interfacial cleanliness. EELS and EDS can be employed to map elemental distributions. (e) STM and STS for atomic-scale surface imaging and spectroscopy to analyze the local density of states and defect characteristics. (f) Qubit energy relaxation measurement (population $P(1)$ versus delay) for extracting the energy relaxation time \(T_1\) and assessing dielectric loss. (g) Qubit spectroscopy for revealing avoided level crossings arising from resonant coupling to a TLS. (h) Time-dependent qubit spectroscopy for tracking TLS frequency fluctuations and spectral diffusion. (i) Device schematic illustrating DC gate biasing and resonator-based microwave characterization for identifying defect-induced loss channels~\cite{Lisenfeld2025Mapping}.}
    \label{fig:Characterization}
\end{figure*}

\subsubsection{Material-Oriented Solutions and Their Characterization}
\label{section:vdW_overview}

Over the past decade, substantial effort has been devoted to identifying material platforms capable of surpassing the performance of aluminum-based transmon qubits. A notable advance has been the widespread adoption of tantalum (Ta) shunt electrodes combined with Al-based JJs, enabling energy relaxation times approaching \( T_1 \sim 1~\mathrm{ms} \)~\cite{place2021new,bland2025millisecond}. This improvement has been largely attributed to the chemically stable surface oxides of Ta, which exhibit significantly lower dielectric loss than those of aluminum or niobium~\cite{bal2024systematic}.

While mitigating losses in the shunt electrodes has yielded impressive gains, attention has increasingly shifted to losses originating in the JJ itself, particularly in the amorphous \(\mathrm{AlO_x}\) tunnel barrier. The strong concentration of electric fields within this amorphous layer leads to enhanced coupling between the transmon and parasitic TLS residing in \(\mathrm{AlO_x}\). These strongly coupled TLS can stochastically drift into resonance with the qubit, producing sudden frequency shifts or energy relaxation ``dropouts'' that degrade device stability and necessitate frequent recalibration of quantum processors~\cite{mohseni2024build}. Consequently, further scaling of superconducting quantum processors will critically depend on strategies that reduce loss and variability associated with the junction itself.

One promising route toward realizing low-loss JJs is the use of two-dimensional (2D) van der Waals (vdW) materials~\cite{novoselov20162d,Geim2013Van,Liu2016Van}. 2D vdW materials embody a materials-by-design paradigm, enabling LEGO-like modular assembly through the vertical stacking of atomically thin layers. In 2D vdW heterostructures, atomically thin crystals are held together by vdW forces rather than by covalent bonds. Compared with conventional amorphous-barrier junctions, 2D vdW heterostructures may offer cleaner and more structurally well-defined interfaces, potentially reducing some of the disordered motifs associated with TLS formation. Because the constituent crystals have atomically abrupt surfaces with few dangling bonds, they can, in principle, be assembled without the lattice-mismatch constraints that often complicate conventional epitaxy~\cite{pizzocchero2016hot}. At the same time, transfer residues, trapped bubbles, edge disorder, and local thickness nonuniformity can still arise depending on fabrication techniques and thus remain important practical concerns.

Furthermore, the Josephson energy can be systematically tuned or engineered by controlling both the thickness of the tunneling barrier and the junction area of the 2D vdW heterostructures, providing a robust means to achieve target qubit frequencies. The broad diversity of available 2D vdW materials, combined with the freedom to engineer stacking sequences and twist angles, offers rich opportunities to implement 2D vdW material-based JJs and qubits. Hence, beyond low-loss JJs, 2D vdW materials and their heterostructures present several compelling advantages for superconducting qubits.

To translate the putative material advantages into genuinely low-loss JJs, the interfaces must be verified to be clean, abrupt, and structurally uniform across multiple length scales. Figure~\ref{fig:Characterization}(a-e) summarizes several characterization techniques for evaluating materials for low-loss JJs, which can be applied to various materials, including 2D van der Waals materials and their heterostructures~\cite{Hegedues2025In,Lisenfeld2025Mapping}. Optical microscopy provides a rapid first screen for cracks, trapped bubbles, transfer residues, and macroscopic nonuniformity, whereas atomic force microscopy (AFM) quantifies thickness, roughness, step morphology, and nanoscale blisters that often signal interfacial contamination. For 2D vdW superconductors and insulators, Raman and photoluminescence (PL) spectroscopy can further provide non-destructive information on layer number, strain, doping, phase purity, and transfer-induced disorder. X-ray photoelectron spectroscopy (XPS) and, where necessary, ultraviolet photoelectron spectroscopy or secondary-ion mass spectrometry are particularly valuable for identifying oxidation states, stoichiometry, adsorbates, and interdiffusion at the JJ interfaces. At higher spatial resolution, cross-sectional scanning transmission electron microscopy (STEM) combined with electron energy-loss spectroscopy (EELS) or energy-dispersive X-ray spectroscopy (EDS) directly reveals barrier-thickness uniformity, lattice order, edge damage, interfacial abruptness, and any residual amorphous layers, while scanning tunneling microscopy/spectroscopy (STM/STS) can resolve atomic-scale defects and the local density of states that may seed subgap states or TLS. In this sense, these tools are not merely descriptive materials probes; rather, they test the central premise of low-loss JJs, namely that the relevant tunneling interfaces can be engineered to avoid the structurally and chemically disordered motifs that dominate dielectric loss in conventional amorphous-barrier junctions.

Once materials are incorporated into JJs or superconducting qubits, additional characterization tools become available to directly evaluate device performance, as shown in Fig.~\ref{fig:Characterization}(f-i). At the JJ level, low-temperature transport measurements, such as I-V characteristics for critical current, normal-state resistance, subgap resistance, excess current, switching-current statistics, and magnetic-field interference patterns, provide an important bridge between ex situ materials analysis and circuit performance by testing the uniformity of tunneling and screening for pinholes, edge-dominated transport, or parasitic dissipative channels. At the qubit level, coherence measurements become the most relevant figure of merit since qubit coherence represents a comprehensive measure of the lifetime of stored quantum information. Although the energy relaxation time \(T_1\) is most commonly reported, the phase coherence time \(T_2\) is equally important because they reflect sensitivity to fluctuating charges, critical-current noise, flux noise, and broader electromagnetic instability in the junction interface and its environment. At the same time, a single coherence time at a single operating frequency should not be overinterpreted, since TLS can be spectrally distributed or localized, and temporally intermittent or fluctuating, so their influence can easily be underestimated or overinterpreted within a limited measurement window. A more practical approach is therefore to map coherence over a wide frequency range, either by explicit flux or gate tuning or by exploiting AC Stark shifts in fixed-frequency qubits~\cite{klimov2018fluctuations,chen2025scalable,lisenfeld2019electric}. When this frequency-resolved characterization is extended in time, it effectively becomes TLS spectroscopy~\cite{klimov2018fluctuations,Carroll2022Dynamics,kim2022effects}, enabling the extraction of TLS spectral density, coupling strength, and switching dynamics, thereby converting qubit coherence from a simple benchmark into a quantitative diagnostic of material quality~\cite{Kim2025Error}.

Beyond a single JJ, TLS-sensitive qubit architectures further amplify microscopic information about loss. Qubits with multiple JJs increase the effective number of participating defects and can therefore enhance sensitivity to weak, distributed dissipation channels. Likewise, fluxonium and related circuits~\cite{manucharyan2009fluxonium,nguyen2019high,nguyen2022blueprint,sun2023characterization,zhuang2025nonmarkovianrelaxationspectroscopyfluxonium,wolff2026structuralcontroltwoleveldefect,azar2026characterizationcomparisonenergyrelaxation}, with their intrinsically broad transition spectra and strong anharmonicity, provide a particularly powerful platform for probing defect dynamics over a wide energy range. For qubits based on low-loss JJs, the key performance target is therefore not simply a high average \(T_1\), but a pronounced suppression of the spectral density of strongly coupled TLS, together with reduced temporal drift, fewer relaxation hot spots, and smaller device-to-device variability. Demonstrating such suppression would provide direct evidence that the material and interface engineering strategy has addressed the dominant microscopic loss channels, thereby establishing these junction platforms as credible building blocks for scalable, low-loss superconducting quantum processors.

\subsubsection{Quasiparticle Tunneling}

\noindent The elementary excitations of a superconductor, known as \emph{quasiparticles}, arise from breaking Cooper pairs in the superconducting condensate~\cite{glazman2021bogoliubov}. Such excitations can be generated by impacts from high-energy particles, which deposit energy exceeding the superconducting gap~\cite{McEwen2022}. If the asymmetry of the superconducting gap across the two electrodes is smaller than the qubit transition energy, a nonequilibrium quasiparticle can absorb energy from the qubit while tunneling across the JJ interface, thereby inducing qubit relaxation~\cite{mcewen2024resisting}. For superconducting quantum processors, impacts from high-energy particles~\cite{Vepsaelaeinen2020} or infrared photons~\cite{diamond2022distinguishing} generate sudden spikes in quasiparticle density that propagate across the processor~\cite{Wilen_2021}, leading to catastrophic correlated errors that are challenging to account for using existing quantum error correction codes~\cite{McEwen2022,mcewen2024resisting}.

Recent studies have highlighted that gap engineering mitigates this process by creating a sufficiently large gap mismatch across the junction, such that quasiparticle tunneling, accompanied by qubit-energy absorption, becomes energetically unfavorable~\cite{Sun2012_qp, Marchegiani2022,mcewen2024resisting,kamenov2024suppressionquasiparticlepoisoningtransmon}. This technique focuses on enhancing the disparity in superconducting gaps across the JJ interface, ensuring that the qubit transition frequency remains below the gap difference across the junction and thus suppresses resonant energy exchange between a quasiparticle and the qubit. In thin superconducting films, where the film thickness approaches the penetration depth, the superconducting gap becomes highly sensitive to variations in film thickness~\cite{Marchegiani2022}. By optimizing the gap difference across the junction by fabricating JJs with electrodes of varying thickness, it is possible to protect against quasiparticle tunneling.

Experimental results have shown significant improvements in qubit coherence under high-energy impact conditions when qubits are designed with substantial gap differences between junction electrodes~\cite{mcewen2024resisting}. This approach paved the way for more effective error reduction in surface-code implementations~\cite{acharya2025}. Future efforts to develop novel JJ materials would benefit from adapting gap engineering to minimize quasiparticle losses and their detrimental effects.

%% file: sections/33_tunability.tex
\subsection{\textit{In Situ} Tunability}
\label{section:tunability}

\begin{figure*}
\centering
\includegraphics[width=17.8cm]{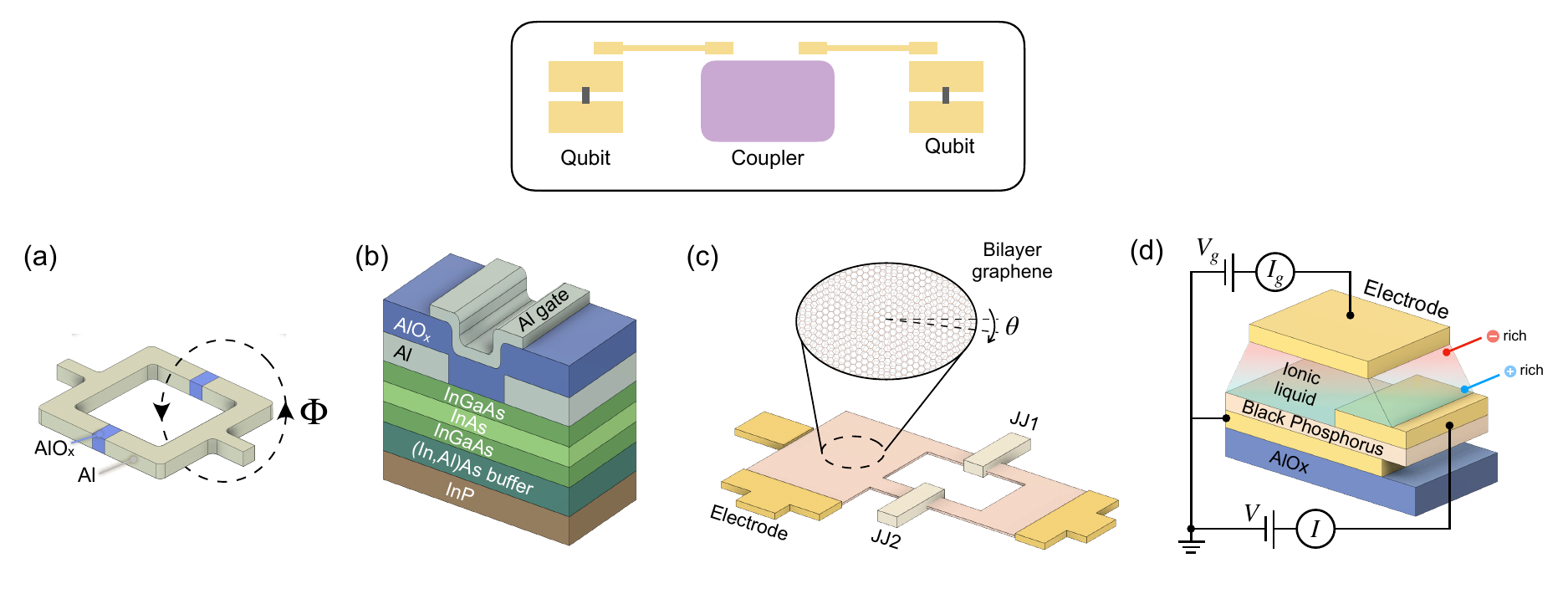}
\caption{\textbf{\textit{In Situ} tunable JJ platforms for superconducting qubits} (a) Al/AlO$_x$/Al SQUID in which magnetic flux $\Phi$ tunes the effective Josephson energy. Beyond flux qubits~\cite{chiorescu2003coherent}, such flux-controlled junctions are commonly employed in tunable couplers to achieve switchable interactions between superconducting qubits. (b) Epitaxial Al/2DEG planar JJ formed on a semiconductor heterostructure, enabling gate control of the critical current through a highly transparent SC/SM interface, making them attractive for scalable gatemon qubits~\cite{casparis2018superconducting,strickland2023superconducting,kjaergaard2017transparent} and voltage-tunable couplers~\cite{Qi2018Controlled,hazard2023superconducting,Casparis2019Voltage,Chen2023Voltage,materise2023tunable}. (c) Monolithic SQUID in magic-angle twisted bilayer graphene with two independently gated weak links, JJ1 and JJ2. Local electrostatic control tunes the critical current in each arm, enabling asymmetric or symmetric SQUID operation and phase-sensitive measurements of the twisted-graphene weak links~\cite{portoles2022tunable}. (d) Ionic-liquid-gated vertical Nb/black-phosphorus/Nb Josephson FET. Electrostatic gating modulates the critical current and junction resistance in an atomic-scale vertical channel, consistent with strongly inhomogeneous, edge- or corner-dominated supercurrent transport in the black-phosphorus barrier~\cite{xu2021vertical}.}
\label{fig:2DvdWJJ}
\end{figure*}

\noindent Tuning the spectrum of superconducting qubits is crucial for flexibility and control, helping alleviate spectral crowding, mitigating frequency collisions with TLSs, and enabling the engineering of entangling interactions between multiple qubits. Tunability in qubits is typically implemented by threading magnetic flux through flux-sensitive qubits, such as split-junction transmons~\cite{koch2007charge}, flux qubits~\cite{chiorescu2003coherent} [Fig.~\ref{fig:2DvdWJJ}(a)], or fluxoniums~\cite{manucharyan2009fluxonium}. By contrast, driving multiple flux-tunable qubits often leads to structural complexity and conflicting requirements. For instance, processor components can form large superconducting loops through the ground plane, thereby increasing sensitivity to flux noise and introducing parasitic coupling and crosstalk. The current required for flux control imposes a significant thermal load on a cryostat, potentially impeding scalability. Furthermore, stray currents from the ground plane are sensitive to wire bonds or air bridges, potentially leading to inaccurate flux biasing or uncontrolled parameter offsets. Therefore, various qubit tuning strategies must be explored to improve the performance and scalability of superconducting quantum processors.

An alternative to flux-tuning is voltage-tuning. In superconducting qubits, the presence of a superconducting island introduces a charge degree of freedom, which allows for voltage control. In practice, this sensitivity is strongly suppressed in the transmon regime. The transmon, for instance, can be made charge-sensitive by reducing the $E_J/E_C$ ratio, gradually mapping to the charge qubit. At the same time, charge-sensitive qubits are highly susceptible to the ubiquitous environmental charge noise, which severely degrades coherence times. On the other hand, charge-sensitive superconducting elements are well-suited for use as couplers, as they mediate interactions between qubits rather than storing quantum information themselves. As a result, their increased sensitivity to charge noise is less detrimental. Moreover, voltage-tunable couplers can avoid the aforementioned issues that arise for flux-tunable circuits.  

Couplers require strong frequency tuning to implement gates while also turning off qubit interactions. More fundamentally, the achievable tuning range in charge-sensitive superconducting elements is ultimately limited by the charging-energy scale $E_\mathrm{C}$, which cannot easily exceed several gigahertz due to stray/coupling capacitances. An alternative to conventional Al/$\mathrm{AlO_x}$/Al JJ-based transmon qubits is the gatemon qubit, in which the Josephson energy is controlled electrostatically through a semiconductor weak link. This enables qubit frequency tuning via gate voltage rather than magnetic flux. Gatemon qubits have been realized using semiconducting nanowire-based JJs~\cite{larsen2015semiconductor, casparis2016gatemon, hays2018direct}, two-dimensional electron gas (2DEG) systems~\cite{casparis2018superconducting,ciaccia2024charge,sagi2024gate,kiyooka2024gatemon,Elfeky2023Quasiparticle}, and 2D vdW JJs~\cite{schmidt2018ballistic,kroll2018magnetic,wang2019coherent,de2021gate,haller2022phase,portoles2022tunable,kononov2020one,kononov2021superconductivity,endres2023current,randle2023gate,jha2025large}. Semiconductor-based gatemon platforms, particularly nanowire-based JJs and 2DEG systems, have demonstrated multi-gigahertz frequency tuning, with coherence times reaching the 20~$\mathrm{\mu s}$ range, whereas implementations of 2D vdW JJs are at an earlier stage. While voltage tunability can, in principle, be used for qubits, its most compelling near-term role is in tunable couplers, where charge-noise sensitivity is less detrimental. For coupler applications, 2DEG systems and 2D vdW JJs are particularly promising, as their planar geometries can relax some of the dimensional constraints that limit the achievable $E_J$ in semiconducting nanowire-based JJs. In the following, we therefore focus on these two material platforms as viable routes toward voltage-tunable superconducting circuits.

\subsubsection{Gate-Tunable 2DEG Systems}

When a semiconductor is coupled to superconducting leads, superconducting correlations are induced into the semiconductor via the proximity effect, enabling dissipationless transport through an otherwise nonsuperconducting material~\cite{Chidambaram2019Microwave, Hays2021Realizing}. Microscopically, the proximity effect originates from Andreev reflection at the SC/SM interface, in which an electron incident from the semiconductor is retroreflected as a hole while a Cooper pair is transferred into the superconductor. As discussed in Section~\ref{section:jj}, coherent Andreev processes across the junction give rise to discrete ABS within the superconducting gap, which carry the Josephson supercurrent (Eq.~\ref{eq:I_ABS}). The ABS energies depend on the superconducting phase difference across the junction and on the number and transparency of the conduction channels. In SC/SM JJs, electrostatic gating provides direct control over channel transparency by changing the carrier density and chemical potential of the semiconductor, thereby enabling in situ tuning of the Josephson current and the effective Josephson energy $E_J$.

2DEG systems are widely used for implementing SC/SM JJs and offer a potentially scalable platform. A representative example is the epitaxial growth of an Al/InGaAs/InAs/InGaAs heterostructure on an InP substrate, as shown in Fig.~\ref{fig:2DvdWJJ}(b). In 2DEG systems, band engineering positions the Fermi level within the conduction band of a few-nanometer-thin InAs quantum well, confining carriers to a 2DEG. This is particularly well-suited for gate-defined JJs due to its high carrier mobility, which enables ballistic transport and channel transparency. Moreover, epitaxial Al/InAs junctions form clean interfaces and highly transparent contacts, a prerequisite for the strong superconducting proximity effect~\cite{Shabani2016Two, strickland2024characterizing}. The quality of SC/SM interfaces in 2DEG systems has been experimentally verified with multiple Andreev reflections~\cite{kjaergaard2017transparent}.

Leveraging these advances, InAs-based gatemon qubits have been realized, exhibiting energy relaxation times of $T_1 \sim 1~\mathrm{\mu s}$ and coherence times of $T_2 \sim 2~\mathrm{\mu s}$~\cite{casparis2018superconducting}. More recently, alternative material platforms have also been explored, including planar Ge-based gatemon qubits, which are still at an early stage and currently exhibit $T_1 \sim 100~\mathrm{ns}$~\cite{sagi2024gate, kiyooka2024gatemon}. Importantly, multiple studies indicate that the semiconductor weak link itself is not the dominant source of decoherence in gatemon devices. Instead, loss mechanisms are primarily attributed to the limited quality factor of the Al film, dielectric loss from thick $\mathrm{AlO_x}$ layers used for top electrostatic gates, and bulk substrate loss originating from the InP substrate. These limitations can be mitigated through materials and architectural optimization, such as increasing the Al thickness to reduce inductive loss, employing low-loss gate dielectrics (e.g., h-BN), or wafer-bonding the heterostructure to a low-loss substrate~\cite{strickland2024characterizing}.

Continued improvements along these directions may enable semiconductor-based gatemon coherence times to exceed $10~\mathrm{\mu s}$. While this remains below the state-of-the-art coherence achieved in transmon qubits, such performance is sufficient to enable gatemon-based tunable couplers~\cite{Qi2018Controlled,hazard2023superconducting,Casparis2019Voltage,Chen2023Voltage,materise2023tunable}. In this architecture, two transmon qubits are coupled via an SC/SM JJ whose transparency is dynamically controlled using voltage pulses, effectively switching the interaction on at finite transparency and off at near-zero transparency. Theoretical studies have suggested that, under favorable assumptions, two-qubit gates may be implemented on $\sim 50~\mathrm{ns}$ timescales with coherent error rates below $0.01\%$~\cite{Qi2018Controlled}.

Further performance gains may be achieved by integrating gatemon-based couplers into a flip-chip architecture~\cite{hazard2023superconducting,sagi2024gate}. In such designs, through-silicon vias (TSVs) can be used to redirect the electromagnetic mode of the gatemon away from the lossy InP substrate and into a low-loss Si interposer, thereby suppressing substrate-induced loss. TSVs additionally provide electromagnetic shielding, reducing crosstalk between circuit elements. Notably, although transmon qubits inherit some decoherence from the coupler, it has been shown that a coupler with $T_2 \sim 1~\mathrm{\mu s}$ can still support transmon coherence times exceeding $100~\mathrm{\mu s}$ while enabling fast, high-fidelity gates~\cite{hazard2023superconducting}.

Beyond tunable couplers, a variety of alternative coupling schemes based on gatemons have been proposed. For example, an SC/SM junction can be embedded in a coplanar waveguide to realize a voltage-controlled superconducting quantum bus~\cite{strickland2023superconducting, Casparis2019Voltage}. Alternatively, the junction may function as a gate-tunable capacitor, enabling coupling between qubits across the transition from a fully transmitting to a fully insulating regime~\cite{materise2023tunable}. Finally, hybrid architectures combining gatemon qubits with transmons enable parametric entangling gates driven by voltage modulation, offering additional flexibility for scalable superconducting quantum processors~\cite{Chen2023Voltage}.

\subsubsection{Gate-Tunable 2D vdW JJs}

\noindent Gate-tunable 2D vdW JJs provide a promising route toward electrically reconfigurable superconducting circuits while inheriting their material advantages, as described in Section~\ref{section:vdW_overview}. Such superconducting qubits may benefit from potentially low-loss, atomically clean interfaces, as well as the large electrostatic tunability of 2D vdW materials and their heterostructures.

Graphene was among the first 2D materials used as a weak link in 2D vdW JJs~\cite{Heersche2007Bipolar,wang2019coherent,aparicio2025gate, schmidt2018ballistic,kroll2018magnetic,haller2022phase}. Similar to SC/SM JJs in 2DEG systems, the supercurrent in graphene-based JJs is driven by ABS. Graphene provides high-transparency conduction channels, owing to its high mobility and ballistic transport, while its low density of states may help reduce the sensitivity of the junction properties to gate-voltage fluctuations. Electrostatic gating further shifts the Fermi energy, thereby altering the carrier density, which can provide direct control over the junction critical current and the Josephson energy $E_\mathrm{J}$. Gate-dependent microwave spectroscopy has provided clear evidence of ballistic transport and high-quality interfaces between graphene and superconducting contacts~\cite{kroll2018magnetic}. Several groups have demonstrated voltage-tunable graphene-based qubits with frequency shifts spanning multiple gigahertz, with the tunability determined by the gate geometry and dielectric thickness~\cite{wang2019coherent,schmidt2018ballistic,aparicio2025gate}. Notably, the qubit transition frequency often exhibits a non-monotonic, oscillatory dependence on gate voltage, resulting in multiple operational sweet spots. These sweet spots may reduce sensitivity to gate-induced charge noise, thereby improving coherence.

Notably, coherent control of graphene-based gatemon qubits has been demonstrated, with reported coherence times on the order of 50~ns~\cite{wang2019coherent}. These relatively short coherence times have been attributed, at least in part, to radiative losses through the gate electrode and fabrication-related imperfections in 2D vdW JJ-based qubits. Radiative losses can be reduced through improved circuit design, for example, by further decoupling the gate electrode from the qubit mode and exploiting symmetry to suppress coupling to the transmon degree of freedom. In addition, integrating low-pass filters on the gate line can attenuate noise at the qubit frequency. It is worth noting that available evidence suggests that the presently observed coherence times are likely influenced substantially by circuit-level loss channels beyond the graphene weak link itself. Future experiments incorporating conventional tunnel-junction transmons or coplanar waveguide resonators on the same chip as graphene-based qubits could help disentangle losses associated with the junction from those arising in shunt capacitors, substrates, or control circuitry. Microwave characterization techniques discussed in Section~\ref{section:vdW_overview} may further provide insight into TLS densities associated with graphene junctions.

Various 2D vdW materials offer complementary approaches for realizing gate-tunable JJs and have demonstrated compatibility with elements required for superconducting circuit integration, including magic-angle twisted bilayer graphene~\cite{de2021gate,portoles2022tunable} [Fig.~\ref{fig:2DvdWJJ}(c)], MoS$_2$~\cite{lee2019two}, MoTe$_2$~\cite{chiu2020flux}, Bi$_2$O$_2$Se~\cite{ying2020magnitude}, and black phosphorus~\cite{xu2021vertical} [Fig.~\ref{fig:2DvdWJJ}(d)]. For example, electrostatic gates can define superconducting leads and non-superconducting junction regions within magic-angle twisted bilayer graphene, thereby enabling the implementation of gate-defined JJs that exhibit both DC and AC Josephson effects while largely avoiding external interfaces caused by multi-material weak links~\cite{de2021gate}. Building on this concept, monolithic SQUIDs based on magic-angle twisted bilayer graphene with two locally tunable weak links have demonstrated phase-coherent interference, symmetric and asymmetric SQUID operation, direct access to the current-phase relation, and large gate-tunable kinetic inductances reaching the $\mathrm{\mu H}$ scale~\cite{portoles2022tunable}. More broadly, twisted graphene superconductors can also serve as circuit elements with unusually large, electrostatically tunable kinetic inductance. For example, twisted trilayer graphene weak links in a SQUID geometry exhibit kinetic inductances up to $\sim 150~\mathrm{nH/\square}$ and coherence lengths approaching $\sim 200~\mathrm{nm}$~\cite{jha2025large}. Few-layer black phosphorus has been used as the vertical weak link in an ionic-liquid-gated Nb/black-phosphorus/Nb Josephson field-effect transistor (FET), in which the critical current and junction resistance are strongly gate tunable despite the ultrashort transport length, consistent with a strongly inhomogeneous supercurrent flow, possibly dominated by edge or corner conduction paths in the BP barrier~\cite{xu2021vertical}. These results show that 2D vdW materials and their heterostructures are not only compatible with superconducting weak-link architectures, but can also support monolithic, phase-coherent, and vertical Josephson devices with strong \textit{in situ} electrical control.

In the context of tunable couplers, 2D vdW JJs can be integrated using strategies similar to those developed for flux-tunable couplers and SC/SM junctions. Because both platforms support gate-tunable transparency and ABS-mediated supercurrents, they can support analogous coupler concepts. However, 2D vdW materials offer gate-tunable superconductivity within a single material, which is distinct from conventional SC/SM JJ-based 2DEG systems. A particularly compelling example is monolayer WTe$_2$, a 2D vdW topological insulator featuring gate-induced superconductivity~\cite{Sajadi2018Gate,Fatemi2018Electrically}. In pristine form, WTe$_2$ exhibits an insulating bulk accompanied by conducting edge modes characteristic of quantum spin Hall insulators~\cite{Moore2010birth}. Upon gating, WTe$_2$ undergoes a transition to a superconducting phase at remarkably low carrier densities, the mechanism for which remains an open question. By patterning multiple gate electrodes across a single WTe$_2$ monolayer, JJs can be defined \emph{in situ}, enabling continuous tuning between fully insulating and fully superconducting regimes. Experimental observations of both superconductivity and the Josephson effect in gated WTe$_2$ have already been reported~\cite{randle2023gate}. Integrating this material platform into superconducting quantum processors could open intriguing opportunities for realizing tunable couplers with exceptionally large on/off ratios, enabled by the ability to electrostatically switch superconductivity itself rather than merely modulating junction transparency.

%% file: sections/34_footprint.tex
\subsection{Device Footprint}
\label{section:footprint_scalability}

\begin{figure*}
\centering
\includegraphics[width=17.8cm]{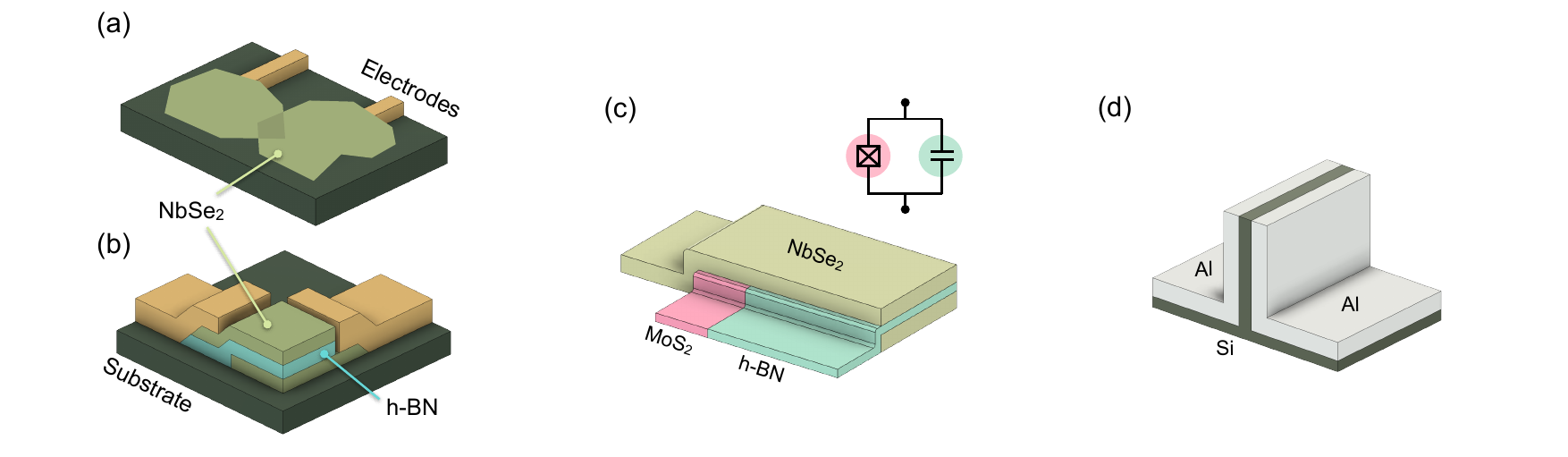}
\caption{\textbf{Compact qubit designs} (a) NbSe$_2$/NbSe$_2$ 2D vdW JJ with a vacuum gap, providing an oxide-free weak link that avoids an amorphous interfacial barrier~\cite{DElia2025coherent,yabuki2016supercurrent,mchugh2023moire}. (b) NbSe$_2$/h-BN/NbSe$_2$ 2D vdW shunt capacitor with a low-loss h-BN dielectric interlayer, offering a promising approach to realize high-coherence superconducting qubits with small layout geometries~\cite{wang2022hexagonal,antony2021miniaturizing}. (c) One promising conceptual design for a highly coherent yet compact 2D vdW transmon qubit architecture, consisting of NbSe$_2$ electrodes, a small-area MoS$_2$ (or WSe$_2$) weak link, and an adjacent h-BN dielectric. (d) A silicon-based fin-shaped merged-element transmon (FinMET) that features high fabrication precision, compactness, and compatibility with CMOS processes~\cite{goswami2022towards}.}
\label{fig:CompactJJ}
\end{figure*}

\noindent Utility-scale, fault-tolerant quantum computers will likely require millions of physical qubits. While the transmon has been foundational for superconducting quantum processors~\cite{koch2007charge,Arute2019supremacy,acharya2025,anferov2024improved}, its large footprint due to the shunt capacitor is a bottleneck to scaling. Typical planar transmons often occupy footprints exceeding $0.1~\mathrm{mm^2}$, which becomes increasingly problematic as chip area, interconnect density, packaging complexity, and cryogenic overhead all tighten with scale~\cite{caldwell2018parametrically,place2021new}. The central challenge is that reducing the qubit footprint generally increases the electric-field participation of the junction barrier and nearby dielectrics. Compact qubit architectures are therefore attractive only insofar as these materials remain sufficiently low-loss and uniform. Therefore, recent efforts aim to miniaturize superconducting qubits while preserving their intrinsic resilience to charge noise, for example, through 2D vdW shunt capacitors or advanced fabrication strategies~\cite{anferov2024improved,wang2022hexagonal,antony2021miniaturizing,zhao2020merged,mamin2021merged,goswami2022towards}.

2D vdW materials naturally enable compact capacitive elements through parallel-plate geometries and, more broadly, a range of compact JJ and qubit platforms~\cite{Balgley2025coherent}. Representative examples include \(\mathrm{NbSe_2/NbSe_2}\)~\cite{DElia2025coherent} [Fig.~\ref{fig:CompactJJ}(a)], \(\mathrm{NbSe_2/WSe_2/NbSe_2}\)~\cite{Balgley2025coherent,balgley2025crystalline}, \(\mathrm{NbSe_2/hBN/NbSe_2}\)~\cite{antony2021miniaturizing, wang2022hexagonal} [Fig.~\ref{fig:CompactJJ}(b)], and hBN-encapsulated graphene JJs~\cite{wang2019coherent}. As shunt dielectrics, 2D vdW insulators such as h-BN can enable compact parallel-plate capacitors with low loss. For example, a monolayer of h-BN is only $\sim$0.3~nm thick, substantially thinner than conventional amorphous dielectric layers used in superconducting circuits, enabling a substantially larger capacitance per unit area, provided that leakage, dielectric strength, and fabrication uniformity can be adequately controlled. Beyond h-BN, high-$\kappa$ 2D vdW dielectrics with ultralow equivalent oxide thickness may offer an additional route to increasing capacitance density in compact junction architectures~\cite{osanloo2021identification}. If such barriers can be integrated while maintaining sufficiently low dielectric loss, they could be attractive for merged-element transmon (\textit{mergemon}) designs~\cite{antony2021miniaturizing,yabuki2016supercurrent,wang2022hexagonal}.

Furthermore, as tunnel barriers or weak-link materials, 2D vdW semiconductors such as MoS$_2$ and WSe$_2$ have been reported to exhibit lower device-to-device variation in the resistance-area (RA) product than conventional amorphous oxides, consistent with improved barrier uniformity and potentially cleaner interfaces~\cite{balgley2025crystalline,lee2019two}. 2D vdW JJs with semiconducting tunnel barriers, for example, a NbSe$_2$/WSe$_2$/NbSe$_2$ heterostructure~\cite{balgley2025crystalline,Balgley2025coherent}, could provide a route toward more compact JJs and, if low-loss interfaces are maintained, improved qubit coherence. Recent MoS$_2$-based implementations have primarily focused on flux-tunable SQUID-type qubits, where coherence may also be influenced by vortex trapping in the junction and readout circuitry~\cite{balgley2025crystalline,lee2019two}. In this context, as shown in Fig.~\ref{fig:CompactJJ}(c), a 2D vdW JJ consisting of NbSe$_2$ electrodes, a small-area MoS$_2$ (or WSe$_2$) weak link, and an adjacent h-BN dielectric may be a promising architecture for a highly coherent yet compact 2D vdW transmon qubit. In this conceptual architecture, the h-BN layer serves as a JJ-embedded shunt dielectric, while the MoS$_2$ (or WSe$_2$) layer serves as the weak-link tunnel barrier, thereby enabling supercurrent tuning.

Another straightforward path to minimizing device footprint is to leverage the intrinsic capacitance of the JJ to eliminate or reduce the need for separate shunt capacitors in quantum circuits. Conventional mergemon qubits employ a thicker oxide grown at higher oxidation pressures together with a larger junction area~\cite{zhao2020merged,mamin2021merged,goswami2022towards}. This approach significantly improves scalability by reducing the footprint by roughly two orders of magnitude while maintaining competitive coherence times of $T_1 \sim 100~\mathrm{\mu s}$ and $T_2 \sim 50~\mathrm{\mu s}$ through optimized thermal annealing processes~\cite{zhao2020merged,mamin2021merged}, which can mitigate the associated junction loss. Other implementations, such as the silicon-based fin-shaped merged-element transmons (FinMETs) shown in Fig.~\ref{fig:CompactJJ}(d), further illustrate this approach through anisotropic etching and float-zone silicon fabrication, which enable high precision and CMOS compatibility~\cite{goswami2022towards}. This compactness, however, comes at the cost of increased electric-field participation in the junction dielectric, so the success of merged-element architectures depends critically on barrier quality, defect density, and processing-induced disorder.

%% file: sections/35_d-wave.tex
\begin{figure*}
    \centering
    \includegraphics[width=17.8cm]{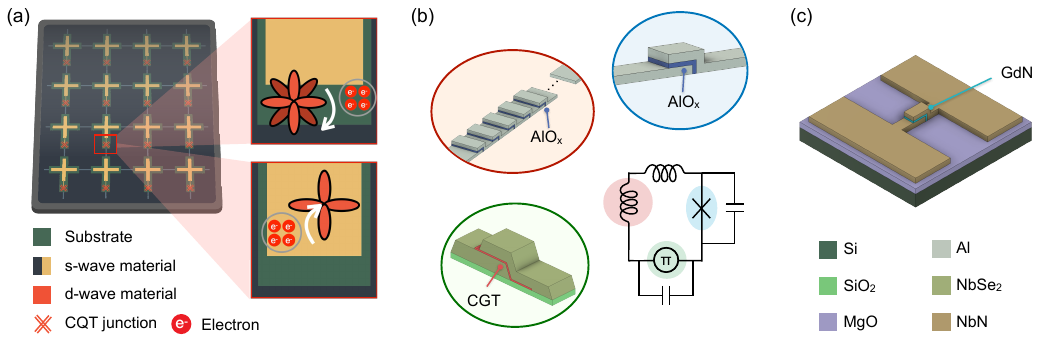}
    \caption{\textbf{Parity protection with CQT junctions.} (a) Array of d-wave Josephson junctions with capacitive shunt for charge noise resilience. Two types of CQT junctions are shown, leveraging either stacking an s-wave SC with a d-wave SC~\cite{patel2024d} (lower) or a relative ${45}^\circ$ rotation between two d-wave SCs ~\cite{brosco2024superconducting, Can2021_natphys_twisted} (upper). (b) Noise-resilient qubit realized in a rhombus circuit geometry. A ferromagnetic insulating Cr$_2$Ge$_2$Te$_6$ (CGT) interlayer realizes a $\pi$-JJ, which can remove the need for external flux biasing. Effective CQT emerges from the interplay between the single junctions and the surrounding Josephson-junction arrays. (c) Intrinsic CQT in a GdN-based $\pi$-JJ. CQT can dominate over single-Cooper-pair tunneling through precise control of the GdN barrier thickness~\cite{Pal2014Pure}.}
    \label{fig:d-wave}
\end{figure*}

\subsection{Noise-Protected Qubits}
\label{section:d-wave}

\noindent The advancement of superconducting quantum processors over the past two decades has been driven primarily by the transmon qubit~\cite{koch2007charge,Arute2019supremacy,acharya2025}. The simple structure of the transmon, involving only a JJ and a large shunting capacitor, eases fabrication, control, and scalability. However, the transmon's weak nonlinearity, due to its simple structure, limits its anharmonicity and does not protect against decoherence. These issues have become more apparent as the margin for error becomes tighter for high-fidelity quantum gates~\cite{Li2023error}. As a result, other superconducting circuits have been explored for enhanced anharmonicity and/or protection, often with increased circuit complexity~\cite{groszkowski2018coherence,nguyen2019high,smith2020superconducting,gyenis2021experimental}. 

One promising approach to protecting quantum information is to leverage Cooper-pair parity~\cite{smith2020superconducting,gladchenko2009superconducting}. In superconducting circuits with an exact or approximate Cooper-pair-parity symmetry, the Hilbert space separates into even- and odd-charge sectors. When the dominant noise operators preserve this parity, matrix elements between the logical states can be strongly suppressed, reducing sensitivity to charge-like decoherence channels. If the junction Hamiltonian is dominated by quartet tunneling rather than single-Cooper-pair tunneling, the two parity sectors can remain nearly decoupled, enabling a protected qubit manifold with enhanced anharmonicity and reduced sensitivity to charge-like decoherence channels. Another envisioned route to hardware-efficient fault tolerance leverages superconducting circuits with dual nonlinearities to host grid states that are intrinsically decoupled from dominant decoherence channels~\cite{gottesman2000encoding,le2019doubly,rymarz2021hardware}. In these devices, parity and redundancy are implemented at the hardware level, providing built-in error correction.

A Cooper-quartet tunneling (CQT) JJ is required to realize a Cooper-pair parity-protected qubit. A CQT junction is effectively described by a dominant second-harmonic Josephson term, corresponding to the coherent transfer of two Cooper pairs. As a result, states of even and odd Cooper-pair numbers cannot be coupled by the tunneling of quartets and are disjoint. In addition, the CQT junction constitutes a key building block of the grid-state qubit and underpins its intrinsic quantum error protection~\cite{nguyen2025superconducting}. The effective CQT junctions have been demonstrated with circuits constructed with SC/SM junctions~\cite{Fukumoriparity} and with Al/Al$\mathrm{O_x}$/Al junctions and superinductors~\cite{Smith2022magnifying,gladchenko2009superconducting}. However, such schemes require external voltage control or magnetic flux, as well as stringent symmetry between constituent elements. Alternatively, several materials that intrinsically favor CQT offer a complementary route in which the desired nonlinearity emerges directly from the junction's physics, without the need for external control parameters. Here, we explore how d-wave SCs and $\pi$-Josephson junctions can be integrated into superconducting processors to realize robust, noise-protected qubits.

In this Subsection~\ref{section:d-wave}, we focus specifically on parity-protected superconducting qubits, where protection arises from the engineered suppression of single-Cooper-pair tunneling and the dominance of higher-order Josephson processes.

\subsubsection{d-wave Superconductors}

\noindent d-wave SC platforms can intrinsically suppress first-order Josephson coupling and thereby favor CQT, depending on material symmetry and twist angle~\cite{patel2024d, brosco2024superconducting, Can2021_natphys_twisted}. CQT in d-wave platforms arises from the symmetry and sign structure of the d-wave order parameter ${\psi_\mathrm{d}}$, which undergoes a sign inversion ${\psi_\mathrm{d}}\rightarrow -{\psi_\mathrm{d}}$ under a ${90}^\circ$ rotation, analogous to the atomic $\mathrm{d}$-orbitals. This symmetry property can be leveraged in two ways. The first is to stack a $\mathrm{d}$-wave SC with an $\mathrm{s}$-wave SC. The Ginzburg-Landau free energy density for this configuration is given by
\begin{equation}\label{eqn:GinzburgLandau_ds}
\begin{split}
    F[\psi_\mathrm{d}, \psi_\mathrm{s}] = &F_\mathrm{d}[\psi_\mathrm{d}] + F_\mathrm{s}[\psi_\mathrm{s}] + A|\psi_\mathrm{d}|^2|\psi_\mathrm{s}|^2 \\
    &+ B(\psi_\mathrm{d}\psi_\mathrm{s}^* + \mathrm{c.c.}) + C(\psi_\mathrm{d}^2\psi_\mathrm{s}^{*2} +\mathrm{c.c.}), 
\end{split}
\end{equation}
where $\psi_\mathrm{d(s)}$ is the order parameter of the $\mathrm{d(s)}$-wave SCs and $F_\mathrm{d(s)}$ is the free energy of each SC. $A$, $B$, and $C$ represent couplings where, in particular, $B$ is the first-order Josephson coupling corresponding to Cooper-pair tunneling, and $C$ is the second-order Josephson coupling corresponding to CQT. If both SCs obey tetragonal symmetry, the free energy density is conserved under a ${90}^\circ$ rotation. However, this rotation leads to ${\psi_\mathrm{d}}\rightarrow -{\psi_\mathrm{d}}$, so the only way for $F[\psi_\mathrm{s}, \psi_\mathrm{d}]$ to be invariant is for $B=0$. On the other hand, the CQT term $C$ is unaffected. Therefore, the first-order Josephson coupling is symmetry-forbidden in the ideal limit, allowing the second-order Josephson term to dominate in a $\mathrm{d/s}$ bilayer stack. While early experiments on $\mathrm{d/s}$ bilayer stacks exist~\cite{sun1994observation, kleiner1996pair, mossle1999c-axis}, evidence for CQT in such structures remains elusive. Advances in heterostructure fabrication and interface control can enable experimental demonstrations of this platform.

Another approach is to stack two d-wave SCs with a relative twist angle in between. The free-energy density is given by 
\begin{equation}\label{eqn:GinzburgLandau_dd}
\begin{split}
    F[\psi_1, \psi_2] = &F_1[\psi_1] + F_2[\psi_2] + A|\psi_1|^2|\psi_2|^2 \\
    &+ B(\psi_1\psi_2^* + \mathrm{c.c.}) + C(\psi_1^2\psi_2^{*2} +\mathrm{c.c.}), 
\end{split}
\end{equation}
where the subscripts denote the top and bottom d-wave SCs. Twisting the second superconducting electrode from ${0}^\circ$ to ${90}^\circ$ results in ${\psi_2}\rightarrow -{\psi_2}$. To conserve the free energy density, the coefficient $B$ must also change sign as the twist angle is increased and crosses a node ($B=0$) when the twist angle is ${45}^\circ$. Therefore, twisted bilayers of d-wave superconductors provide a natural route toward CQT-dominated Josephson coupling. Experimental signatures consistent with dominant CQT or second-harmonic Josephson coupling have been reported in 2D vdW twisted bilayer BSCCO, a high-temperature d-wave SC~\cite{Zhao2023bscco}.

The next milestone is to shunt a $\mathrm{d/s}$ or $\mathrm{d/d}$ junction with a large capacitor, as shown in Fig.~\ref{fig:d-wave}(a), to realize a protected superconducting qubit~\cite{brosco2024superconducting, patel2024d}. Like the transmon, the large capacitive shunt suppresses charging noise and enhances phase coherence. Three criteria must be met for a successful implementation. A pristine, high-transparency interface is required to maximize the desired quartet-tunneling amplitude and minimize uncontrolled single-pair processes. Independently, a sufficiently large shunt capacitance is needed to suppress charge dispersion and stabilize the protected qubit manifold.

Second, precision in fabrication is required to suppress the first-order Josephson coupling. Cooper-pair tunneling lifts the degeneracy of the logical states and reduces the anharmonicity and protection of the qubit. While a small degree is tolerable and can even help reduce charge noise sensitivity, the protection and anharmonicity are lost once the first- and second-order Josephson couplings become comparable. Therefore, precise control of the twist angle will be important when utilizing a twisted bilayer. 

Third, quasiparticle tunneling must be suppressed. Nodal quasiparticles survive in the $\mathrm{d}$-wave order parameter down to low energies. When quasiparticles tunnel across a junction, they absorb and depolarize the qubit energy. Fortunately, the energy barrier in the double-well potential of the $\mathrm{d/s}$ junction and the momentum mismatch due to the twisting in $\mathrm{d/d}$ junctions can strongly suppress quasiparticle tunneling~\cite{patel2024d,brosco2024superconducting}. This resembles the gap engineering discussed in Section~\ref{section:tls}. 

Integrating d-wave SCs into superconducting quantum processors presents an exciting and complex challenge for materials science. Despite their potential, d/s SC bilayers remain largely unexplored, with key issues such as lattice matching, interfacial coherence, and compatible growth temperatures requiring careful consideration~\cite{patel2024d}. For twisted bilayer structures, the mechanical exfoliation and transfer method currently represents the state-of-the-art fabrication method, but it is inherently handcraft-intensive, exhibits significant variability across fabrication rounds, and is not scalable at this stage, making it only suited for proof-of-concept demonstrations with single or two qubits. Achieving scalable integration will ultimately depend on developing CMOS-compatible, wafer-scale, reliable growth techniques for emerging materials, including high-quality d-wave SCs. Prospective strategies are outlined in Subsection~\ref{section:epitaxial}.

\subsubsection{$\pi$-Josephson Junctions}

\noindent $\pi$-Josephson junctions ($\pi$-JJs), characterized by an intrinsic 
 $\pi$-shift in their current-phase relation (CPR), present a compelling pathway toward scalable, intrinsically protected qubits. In contrast to conventional Cooper pairing, where electrons pair with opposite spins and momenta, a ferromagnetic material can introduce a magnetic exchange energy that lifts spin degeneracy. To compensate, each electron in the Cooper pair acquires a finite additional momentum, resulting in a net center-of-mass momentum for the pair. This momentum leads to spatial oscillations of the superconducting order parameter within the ferromagnet interlayer. Thus, the phase shift arises naturally in systems that break time reversal symmetry, such as SC/Ferromagnet/SC (SFS) JJs~\cite{Golubov2004current, Ryazanov2001Coupling, Larkin1964nonuniform, Fulde1964Superconductivity}. If inversion symmetry is preserved, the CPR constrains the equilibrium phase difference to either $0$ or $\pi$, depending on the sign of the effective Josephson coupling. Thus, by precisely tuning the ferromagnetic interlayer thickness, an intrinsic $\pi$-phase shift can be engineered, enabling the realization of $\pi$-junctions with tailored quantum properties.

Since Ryazanov’s seminal demonstration of the $\pi$-Josephson effect in a CuNi alloy over two decades ago, a diverse range of $\pi$-JJs has been realized using pure ferromagnetic elements, alloys, ferromagnetic insulators, and 2D vdW ferromagnets~\cite{Ryazanov2001Coupling,Birge2024Ferromagnetic}. However, only a select few exhibit properties suitable for quantum applications. First, the characteristic decay/oscillation length of superconducting pair correlations in the ferromagnet, which governs the transition between the $0$ and $\pi$ states, must be long enough ($\sim$1~nm) to be experimentally accessible, yet short enough to sustain a large critical current. Strong ferromagnets, with their large exchange splitting, suppress ferromagnetic coherence, making them unsuitable. Second, the magnetic barrier must be both uniform and magnetically soft. For instance, while Ni can carry supercurrent effectively, its magnetic hardness requires large initialization fields and can lead to domain formation, limiting junction size and critical current. Third, dissipation must be minimized by ensuring the Stewart-McCumber parameter, $\beta_c = (2e/\hbar)I_cR_N^2C$, is larger than 1, which corresponds to an underdamped regime. In superconducting circuits incorporating SFS junctions as passive phase shifters, the Caldeira-Leggett model predicts that loss is minimized when the junction is underdamped~\cite{Kato2007Decoherence}. This may help explain the limited coherence ($\sim$1~$\mu$s) observed in superconducting qubits integrating overdamped SFS junctions~\cite{Kim2024Superconducting}. Therefore, for practical quantum applications, the ferromagnetic barrier must be underdamped, magnetically uniform, and capable of sustaining a sufficiently large critical current density.

Ferromagnetic insulators are especially promising for $\pi$-JJs because their insulating character can suppress quasiparticle transport while still enabling a $\pi$-shifted Josephson coupling~\cite{Kawabata2006Macroscopic}. In particular, 2D vdW ferromagnets provide atomically sharp interfaces and precise layer-by-layer control~\cite{ai2021van,kang2022van,idzuchi2021unconventional}. $\pi$-JJs based on 2D vdW materials have been demonstrated using $\mathrm{NbSe_2}$/$\mathrm{Cr_2Ge_2Te_6}$/$\mathrm{NbSe_2}$ heterostructures, where $\mathrm{NbSe_2}$ serves as the superconducting electrode of choice due to its compatibility with 2D vdW materials and its resilience to high in-plane magnetic fields (30–40 T) without flux trapping. A clear $0$-$\pi$ transition is observed at a barrier thickness of 8.4 $\mathrm{nm}$, corresponding to a critical current density of 40 $\mathrm{nA/\mu m^2}$ ~\cite{kang2022van}. However, this low critical current density underscores the need for alternative ferromagnetic-insulator-based $\pi$-junctions. Promising candidates include sputtered $\mathrm{NbN}$/$\mathrm{GdN}$/$\mathrm{NbN}$ junctions, which exhibit $0$-$\pi$ transitions. More broadly, recently explored 2D vdW ferromagnetic metals such as Fe$_3$GeTe$_2$ may provide additional routes toward engineered $\pi$-junction behavior, although their suitability as low-loss barriers remains unexplored~\cite{Senapati2011Spin,hu2025proximity}. Below, we elaborate on strategies for implementing superconducting qubits based on $\pi$-JJs.

Before full integration in a complex circuit, preliminary experiments should validate the performance of a candidate $\pi$-JJ. One approach is to shunt the $\pi$-JJ with large superconducting electrodes, forming a transmon-like circuit. This enables the extraction of its Josephson energy, which can be compared with current–voltage measurements and optimized by adjusting the junction size and barrier width. High-throughput characterization is possible by fabricating multiple junctions with varying dimensions on a single chip. Once optimal parameters are identified, the $\pi$-JJ can be placed in parallel with a 0-JJ to form a SQUID—both confirming its intrinsic phase shift and benchmarking the symmetry between the zero and $\pi$ junctions.

We propose two conceptually distinct architectures. The first uses a $\pi$-JJ as a passive phase shifter within a circuit-engineered effective CQT element, whereas the second aims to exploit a junction whose intrinsic CPR is already dominated by the second harmonic. In the first approach, shown in Fig.~\ref{fig:d-wave}(b), a superconducting circuit can be designed in a rhombus geometry consisting of a 0-JJ (such as a tunnel junction), a $\pi$-JJ, and superconducting wires. This setup resembles the effective CQT junction realized with Al/$\mathrm{AlO_x}$/Al junctions, where the superinductors enable higher-order Cooper-pair tunneling~\cite{Smith2022magnifying,gladchenko2009superconducting}. However, using a $\pi$-JJ eliminates the need for external magnetic flux control, reducing noise from control lines and enhancing scalability by decreasing the number of control parameters. A key challenge in this architecture is ensuring that the Josephson energies of the 0- and $\pi$-JJs are precisely matched, as any asymmetry can reintroduce Cooper-pair tunneling, thereby weakening parity protection. In the second approach, shown in Fig.~\ref{fig:d-wave}(c), symmetry constraints can be circumvented by directly leveraging the intrinsic CQT of the $\pi$-JJ, providing a conceptually simpler route toward parity-protected qubits. Second-harmonic contributions in the CPR have been observed in $\mathrm{NbN}$/$\mathrm{GdN}$/$\mathrm{NbN}$ junctions, suggesting a promising route toward achieving CQT-dominated behavior~\cite{Pal2014Pure}. Near the precise $0$-$\pi$ transition point, the first-harmonic Josephson term can be strongly suppressed, allowing the second harmonic to dominate. Additionally, shunting the junction with a large capacitor can provide parity protection, similar to the d-wave qubit discussed earlier. Achieving this goal will require precise control over the barrier thickness and the realization of a sufficiently large CQT amplitude. Efforts in this direction will pave the way for $\pi$-JJ-based qubits to become a viable platform for robust, scalable quantum computation, as the required nonlinearity can emerge directly from junction physics rather than from circuit elements, providing a promising route toward materials-enabled protected superconducting qubits.

%% file: sections/60_fabrication.tex
\section{Challenges \& Opportunities in Nanofabrication}
\label{section:fab}

\noindent The remarkable progress of the CMOS IC chip industry has been driven by coordinated advances in lithography, deposition, etching, process integration, and packaging~\cite{shin20252d,yoon2025enabling,jung2026advances}. Superconducting quantum technologies will likewise require a multidisciplinary nanofabrication strategy, but with an added emphasis on ultra-low-loss materials, pristine interfaces, and cryogenic reliability.

A key fabrication challenge is precisely depositing superconducting thin films with minimal defect density while maintaining high interface quality. Josephson junction patterning and device architecture optimization must also ensure low-loss interfaces and reproducible junction characteristics. Moreover, as quantum devices operate in cryogenic environments, ensuring long-term reliability under extreme conditions is essential. Addressing these challenges is crucial to achieving scalable, high-performance, industrially viable quantum devices.

This section examines the state-of-the-art fabrication techniques critical to JJ-based quantum technologies and explores their scalability and integration potential. Multi-angle evaporation, a widely used method in superconducting circuit fabrication, is analyzed in Subsection~\ref{section:evap}, where we discuss its role in JJ formation, its impact on junction uniformity, and potential improvements.
Subsection~\ref{section:CMOS} explores the transition from university-scale fabrication to foundry-compatible CMOS processes, addressing key process adaptations required to bridge the gap between traditional conventional CMOS IC chip manufacturing and quantum circuit fabrication. In Subsection~\ref{section:epitaxial}, we introduce the integration of novel materials via mechanical exfoliation for the proof-of-concept of the materials and discuss directions for interface control and large-scale material synthesis.

%% file: sections/61_evap.tex
\subsection{Conventional Multi-Angle Evaporation for JJ}
\label{section:evap}

\noindent Since the 1970s, the fabrication of Josephson tunneling junctions has relied on the multi-angle evaporation of superconducting electrodes with an insulating layer positioned between~\cite{dolan1977offset}. Recent decades have seen remarkable advancements in this method, accelerating the development of superconducting quantum computing technologies. These advances have been driven by dramatic improvements in material quality and interfaces, alongside enhanced yields enabled by robust lithography techniques. Leveraging these developments, many academic and industrial groups have adopted a three-stage manufacturing procedure, which now routinely achieves coherence times in the range of hundreds of microseconds~\cite{kreikebaum2020improving,Osman2021Simplified,muthusubramanian2024wafer}.

Photolithography is the first stage of the manufacturing process and is used to fabricate the linear microwave components of the quantum processor. This begins with thoroughly treating the substrate to ensure a pristine interface that minimizes TLS. Common substrate choices include sapphire, valued for its low-loss properties but challenging to process due to its hardness, and high-resistivity silicon, which benefits from a well-established industry but suffers from more significant surface and bulk losses. The first treatment is a Piranha solution (a mixture of sulfuric acid and hydrogen peroxide) to remove organic residue~\cite{place2021new}. Additionally, for Si substrates, a mixture of ammonium fluoride and hydrogen fluoride, \textit{i.e.}, buffered oxide etchant (BOE), is used to remove lossy surface SiO$_2$~\cite{biznarova2024mitigation}. Minimizing the transfer time from substrate cleaning to the deposition chamber is also essential to prevent recontamination and the regrowth of ambient oxides~\cite{fritz2019optimization}.

After cleaning, the SC is deposited onto the substrate. Metals such as Al and Nb can be sputtered directly. On the other hand, Ta, which has recently been favored for its low-loss surface oxide, requires either heating the substrate to $\sim 500^\circ \mathrm{C}$ or depositing a thin Nb seed layer to promote the growth of the $\alpha$-Ta phase, which offers a superior critical temperature and lower loss compared to $\beta$-Ta~\cite{place2021new, Urade2024Microwave}. Following deposition, photoresist is spin-coated, and microwave components are patterned using a near-ultraviolet laser. With a resolution of $\sim \mathrm{1~\mu m}$, this method suits linear components ($>\mathrm{10~\mu m}$) and offers a fast write time ($\lesssim$ 100~min for a 300~mm wafer) compared to slower, high-precision techniques like electron-beam (e-beam) lithography. Finally, the exposed resist is developed, and the unwanted metal is etched away, revealing the defined processor components.

In the second stage, the JJs are fabricated using e-beam lithography. Before Al deposition, the substrate's ambient oxides must be removed. This can be achieved using BOE for Nb and Ta. For Al, this is not possible since Al is etched by BOE~\cite{altoe2022localization,crowley2023disentangling}. Alternatively, ion milling in the junction-deposition chamber can be used instead of wet etching. However, special care must be taken to avoid excessive surface roughness due to the bombardment of ions~\cite{Yost2020Solid,VanDamme2023Argon,Mergenthaler2021Effects}.

After treatment, the junctions are patterned onto a spin-coated bilayer of e-beam resist, which enables the formation of bridges and overhangs required for shadow evaporation. This step is crucial for meeting the yield and uniformity requirements outlined in Subsection~\ref{section:yield}. There are two main techniques for defining junctions. The Dolan style uses a suspended resist bridge to shadow the evaporated aluminum, with junction dimensions determined by the bridge width/length, evaporation angle, and resist height~\cite{dolan1977offset}. This method, which is essential for high-inductance superconducting circuits, is sensitive to variations in resist height~\cite{nguyen2019high}. In contrast, the Manhattan style relies on a cross-shaped trench that precisely defines the junction size, making it resilient to resist variations~\cite{Potts2001Novel}. However, the orthogonal geometry complicates the fabrication of junction arrays. The choice of evaporation technique ultimately depends on the specific superconducting circuit and application.

Regardless of the lithographic style, a uniform, defect-free resist coating is essential for accurately defining the junctions. This requires a consistent ambient pressure during spin coating to avoid defects~\cite{kreikebaum2020improving}. For e-beam lithography, proximity effect correction is applied to account for electron backscattering. Additional care is needed for nonconductive substrates, such as sapphire, to prevent charge buildup, which can be mitigated by depositing a thin metal layer on top of the coated resist~\cite{place2021new, Aassime2013Anti, Zheng2023Fabrication}. Finally, the developer’s temperature, duration, and sonication must be carefully optimized to avoid under- or overdevelopment and to preserve the integrity of overhanging patterns~\cite{kreikebaum2020improving,Osman2021Simplified,muthusubramanian2024wafer}.

The junctions are finalized using double-angle evaporation of Al, with \textit{in situ} oxidation between the depositions. Al is the electrode of choice for evaporation due to its relatively low melting point compared to Nb or Ta, which require sputtering. Furthermore, the thin, self-limiting Al oxide is suitable as a tunnel barrier, as its thickness determines the Josephson energy. The oxidation pressure and duration must be carefully calibrated to achieve the desired qubit parameters. Dynamic oxidation, which maintains constant pressure by modulating oxygen flow, offers improved reproducibility and uniformity compared to static oxidation, where a fixed initial amount of oxygen is introduced~\cite{osman2023mitigation, kreikebaum2020improving,fritz2019structural,chen2023optimization}. However, some nonuniformity is unavoidable due to the finite distance between the sample and the Al source in the evaporation chamber, causing radial variations in the evaporation angle~\cite{kreikebaum2020improving, osman2023mitigation}. To mitigate this, the junction dimensions can be pre-adjusted to compensate for radial variability. 

In addition to the oxidation and lithographic parameters, the thickness of the deposited electrodes plays a critical role in qubit performance. By varying the thickness between the junction electrodes, quasiparticle tunneling across the junction can be suppressed (Subsection~\ref{section:tls}). In superconducting thin films, the film thickness directly influences the superconducting gap. By engineering a gap difference larger than the qubit energy, quasiparticle tunneling, and thus qubit decay can effectively be suppressed. A gap difference exceeding $h\times\mathrm{10~GHz}$, which is larger than typical qubit frequencies, can be achieved, for example, with a bottom electrode thickness of $10~\mathrm{nm}$ and a top electrode thickness of 100~nm. After depositing the top electrode, the junctions are defined by lifting off the resist and excess metal using a solvent.

In the final stage, a patch layer is deposited to establish a reliable electrical contact between the first two stages~\cite{Gruenhaupt2017argon,Osman2021Simplified,bilmes2021situ}. The lithography process mirrors the second stage, but without any prior surface treatments. The patches are patterned at the intersections of the base layer and junction electrodes. Ion-milling removes ambient oxides in the patch region, followed by Al evaporation to form a strong electrical connection. After lift-off, the fabrication of the superconducting quantum processor is complete. The complete assembly, containing dozens of processor dies, can be probed at room temperature to sieve for defects. Those that meet operating standards are packaged into a printed circuit board box and loaded into the cryostat.

The three-stage process can be adapted to improve processor scalability. Achieving greater scalability involves transitioning from planar geometries to out-of-plane interconnects~\cite{Rosenberg2020Solid}. Multi-chip architectures separate quantum components from control components, placing them on two distinct chips connected via flip-chip bump bonding. An optional third chip, featuring metalized TSV, can be placed between the chips to enhance isolation~\cite{Rosenberg20173D, Yost2020Solid}. The control chip is typically fabricated using photolithography, while the quantum chip follows a three-stage process. Flip-chip bump bonding requires specialized tools and techniques to deposit indium pillars and precisely align the chips~\cite{Kosen2022Building}. Although this adds complexity, it offers significant advantages, including efficient signal routing, reduced crosstalk, and independent optimization of surface treatment and passivation~\cite{Kosen2024Signal,Zheng2022Nitrogen,bal2024systematic,Alghadeer2022Surface}. Alternatively, out-of-plane interconnections can be achieved using spring-loaded pogo pins, machined structures, or adhesives~\cite{Spring2021High,Rahamim2017Double,Patterson2019Calibration,Conner2021Superconducting,Bronn2018High}, providing a versatile pathway to scalability.

Finally, while shadow-evaporated JJs have advanced remarkably, inherent limitations may hinder the progress of superconducting quantum technologies. The primary challenge is that metal deposition occurs after resist patterning, leading to three key issues. First, the pristine substrate-metal interface is disrupted (or undone) by air exposure during lithography. Second, materials that require high-temperature deposition are incompatible since heat destabilizes and deforms organic resist. Third, dangling resist polymers can contaminate the electrodes and the junction barrier, contributing to junction aging and performance degradation over time. In addition to these issues, angled evaporation is sensitive to resist height, which reduces junction uniformity. These challenges have spurred growing interest in etch-based fabrication of JJs, which will be explored in Subsection~\ref{section:CMOS}.

%% file: sections/62_300mm.tex
\subsection{Towards Superconducting Quantum Foundries}
\label{section:CMOS}

\noindent As quantum technology evolves beyond its early exploratory phase toward scalable, high-quality devices, transitioning from bespoke academic cleanroom processes to foundry-compatible, wafer-scale fabrication will be essential. This transition marks a pivotal shift in quantum device manufacturing that mirrors the trajectory of the SM IC chip industry, progressing from academic experimentation to industrial-scale production of silicon microchips with nanoscale features.  

To date, the fabrication of JJ-based quantum devices has primarily occurred in academic settings, emphasizing rapid turnaround and iterative development using processes tailored to specific experiments. In contrast, SM foundries rely on highly standardized, high-throughput workflows to produce chips with ultra-fine ($\sim$10 nm) CMOS transistors at high densities (up to $\sim$100 billion devices per chip). A comparison of representative process paradigms is provided in Table~\ref{cmos_compare}. As the superconducting quantum community begins to adopt the high-precision infrastructure of conventional silicon foundries, the central challenge is to develop process flows that maintain ultra-low-loss materials and interfaces, which are crucial for high-coherence devices while scaling to industrial levels of reproducibility and yield.







\begin{table*}[t!]
    \centering
    \caption{Comparison of prototype-oriented fabrication and foundry-compatible manufacturing for JJs, qubits, and superconducting circuits}
    \label{cmos_compare}
    \resizebox{0.98\linewidth}{!}{%
    \renewcommand{\arraystretch}{1.35}
    \begin{tabular}{>{\raggedright\arraybackslash}p{0.22\textwidth}
                    >{\raggedright\arraybackslash}p{0.36\textwidth}
                    >{\raggedright\arraybackslash}p{0.36\textwidth}}
    \hline
    \hline
    \textbf{Category} & \textbf{Prototype-oriented fabrication} & \textbf{Foundry-compatible manufacturing} \\
    \hline\hline

    Primary objective 
    & Rapid iteration, flexible prototyping, and materials exploration 
    & Wafer-scale uniformity, reproducibility, yield, and process integration \\

    Typical lithography 
    & EBL for junctions; optical lithography or direct-write lithography for larger circuit features 
    & DUV photolithography or other stepper/scanner-based lithography for wafer-scale patterning; EBL used mainly for development or niche features \\

    Junction definition 
    & Shadow evaporation with Dolan/Manhattan resist structures; lift-off-based JJ formation 
    & Subtractive or semi-subtractive processes, including trilayer and layer-by-layer JJ flows \\

    Tunnel barrier formation 
    & \textit{In situ} oxidation during double-angle evaporation; barrier thickness set by oxidation time/pressure 
    & Predeposited trilayer barriers or layer-by-layer oxide growth after controlled surface cleaning and oxidation \\

    Interlayer cleaning and oxide removal 
    & Wet etch or \textit{in situ} ion milling immediately before deposition 
    & Integrated surface preparation modules with controlled native-oxide removal and tighter process monitoring \\

    Metallization 
    & Predominantly evaporation for Al-based JJs; sputtering used for base films and some superconducting layers 
    & Sputtering, ALD, PVD, and other wafer-compatible thin-film tools with tighter thickness and across-wafer uniformity control \\

    Pattern transfer 
    & Lift-off is common; etching used selectively 
    & Dry etching, planarization, spacer-defined isolation, and additive/subtractive routing designed for scalable integration \\

    Dielectric isolation and planarization 
    & Often minimized or avoided to reduce dielectric loss; limited use of temporary process dielectrics 
    & More extensive use of spacers, planarization, and temporary isolation layers, which must be engineered to avoid added loss \\

    Metrology and process control 
    & Device-level validation, room-temperature screening, and iterative tuning of local process conditions 
    & Wafer-scale process control, inline metrology, across-wafer critical-current uniformity, defect monitoring, and statistical yield analysis \\

    Main strength 
    & Fast development cycle and compatibility with unconventional device concepts and novel materials 
    & High reproducibility, tighter parameter spread, backend compatibility, and scalable manufacturing \\

    Main limitation 
    & Limited throughput, weaker wafer-scale uniformity, and reduced compatibility with industrial integration 
    & Higher process complexity and stricter requirements for loss control, materials compatibility, and integration of nonstandard junction materials \\

    \hline
    \hline
    \end{tabular}
    }
\end{table*}

One of the most significant challenges in this transition lies in fabricating high-quality JJs, which differ markedly from standard SM foundry processes. To achieve the typical JJ feature size of $\sim$100~nm, electron-beam lithography (EBL) has traditionally been employed because it can define features as small as $\sim$7~nm. EBL is maskless, offers flexibility, and enables rapid iteration, making it well-suited for prototyping and exploring new device architectures. However, EBL is inherently time-consuming, prone to errors such as charging effects and electron scattering, and, crucially, not fundamentally required for JJ fabrication at the $\sim$100~nm scale when DUV-compatible subtractive flows are available.

Notably, the required feature sizes for JJs are within the resolution limits of deep ultraviolet (DUV) photolithography. Consequently, there has been growing interest in adapting foundry-grade DUV tools for JJ patterning. Emerging approaches generally fall into two categories: (1) tri-layer junction fabrication on wafers with a uniform junction material stack~\cite{anferov2024improved,tolpygo2015}, and (2) layer-by-layer patterning strategies incorporating \textit{in situ} native oxide removal and controlled oxide growth~\cite{ke2025scaffold,van2024advanced}. We examine both approaches in detail, highlighting their advantages and limitations.

\begingroup
\setlength{\dbltextfloatsep}{8pt}
\setlength{\abovecaptionskip}{4pt}
\setlength{\belowcaptionskip}{0pt}

\begin{figure*}[!t] 
    \centering
    \includegraphics[width=17.8cm]{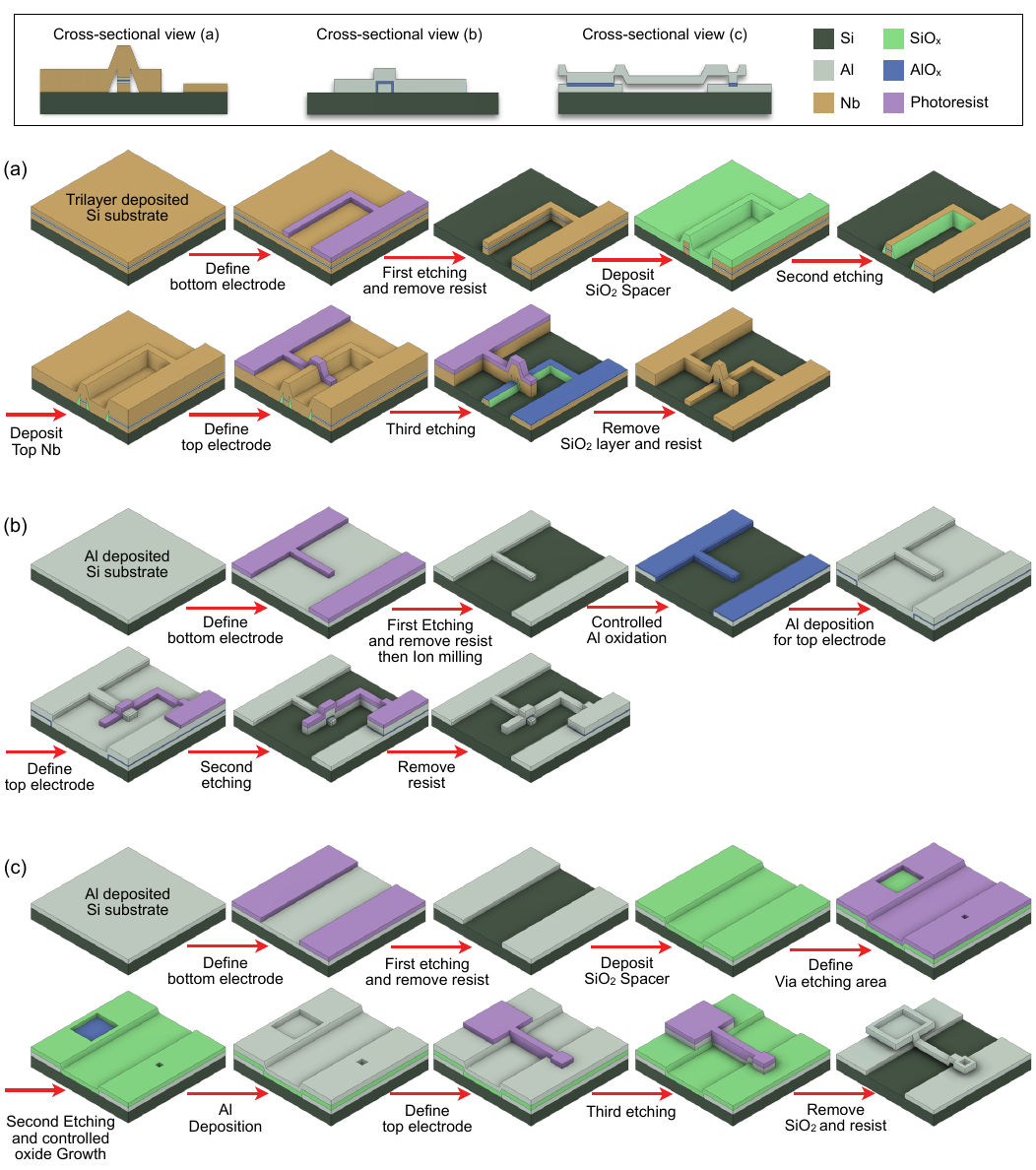}
    \caption{{\textbf{Foundry-compatible JJ fabrication processes at the wafer scale.} (a) Nb/Al-$\mathrm{AlO_x}$/Nb trilayer junctions. Multilayers are deposited and patterned using SiO$_2$ spacers to define junctions and avoid shorts, followed by top Nb wiring and spacer removal~\cite{anferov2024improved}. (b) Al/$\mathrm{AlO_x}$/Al JJs defined by etching. The bottom Al is patterned, ion-milled, oxidized, and topped with a second Al layer to form the junction~\cite{van2024advanced}. (c) Planarized Al/$\mathrm{AlO_x}$/Al JJs. The bottom Al and circuits are patterned first. Then the SiO$_2$ spacer forms via windows and controlled oxidation creates the barrier before the final top Al deposition and spacer removal~\cite{ke2025scaffold}.}} 
    \label{cmos:fig:junctionStep} 
\end{figure*} 
\endgroup

\subsubsection{Tri-Layer Process}
\noindent In the first approach, as shown in Fig.~\ref{cmos:fig:junctionStep}(a), all superconducting and insulating layers forming the JJ are deposited before any lithographic steps. These materials can be grown with atomic-level precision in ultra-high vacuum using advanced thin-film deposition techniques. Because the full trilayer stack is deposited before lithography, barrier thickness and electrode geometry can be controlled more uniformly across the wafer. This improves the reproducibility of the junction's critical current density, even though the tunnel barrier may remain amorphous, depending on the chosen material system. Most importantly, variations in barrier thickness are suppressed, which significantly improves the uniformity and consistency of JJs at the wafer scale, as the critical current density of a JJ depends exponentially on its barrier layer thickness. DUV lithography and anisotropic dry metal etching enable the fabrication of JJs with high geometric area precision. The precise subtractive process further improves junction uniformity and consistency over a wafer compared to the liftoff processes, as a JJ's critical current is proportional to its area. 

While deposition, DUV lithography, and anisotropic etching enable the fabrication of JJs with exceptional uniformity and wafer-scale consistency, a critical challenge lies in routing current between the bottom and top electrodes without introducing electrical shorts. This step is essential for integrating the junctions into the broader quantum circuit and achieving a reliable junction yield across the wafer. Early tri-layer fabrication approaches~\cite{tolpygo2015, macklin2015near} addressed this issue by permanently encapsulating the exposed superconducting leads with a lossy dielectric isolation layer. However, this strategy increased dielectric loss, resulting in significantly lower coherence than Al-based junctions fabricated via angled evaporation.

A recent advancement overcame this limitation by introducing a removable isolation layer~\cite{anferov2024improved}. In this approach, a 
$\mathrm{SiO_2}$ layer is temporarily deposited along the junction electrodes' exposed surfaces through an additional deposition sequence, lithography, and etching. This layer provides physical isolation, preventing electrical contact between the existing bottom electrode and the subsequently deposited top wiring layer. Crucially, the isolation material can later be selectively removed, thereby mitigating additional dielectric loss while preserving reliable junction connectivity, offering a scalable, low-loss solution compatible with advanced quantum circuit architectures.

A critical step in the process outlined in Fig.~\ref{cmos:fig:junctionStep}(a) involves the selective removal of the temporary SiO$_2$ isolation layer—a delicate procedure requiring careful experimental optimization. Because the SiO$_2$ spacer resides beneath the metal wiring, conventional top-down anisotropic physical etching proves ineffective. Conversely, aggressive isotropic chemical etching risks damaging the underlying junction structure if not correctly optimized for material selectivity. Incomplete removal can also degrade coherence due to residual dielectric loss associated with the amorphous SiO$_2$.

Recent work using Nb/Al-$\mathrm{AlO_x}$/Nb trilayer junctions with a temporary SiO$_2$ spacer demonstrated improved coherence times approaching $\mathrm{60~\mu s}$, achieved via an optimized selective wet chemical etching protocol~\cite{anferov2024improved}. However, finite selectivity and surface tension effects limit this wet etching process, particularly when removing SiO$_2$ from confined geometries. Highly selective vapor-phase hydrofluoric acid (VHF) etching offers a promising alternative, allowing conformal removal of SiO$_2$ with reduced risk to the surrounding superconducting structures.

The overall fabrication process proceeds as follows: SCs and insulators are first deposited on a silicon wafer, followed by photolithographic patterning and etching to define the junction electrodes. Then, a SiO$_2$ spacer is added and patterned to isolate the bottom and top electrodes during subsequent electrical processing. The top Nb wiring layer is deposited and etched to connect the JJ's upper electrode to the planar circuit. Finally, the SiO$_2$ spacer is selectively removed, completing the junction without introducing lossy dielectric interfaces. With further refinement, this process holds promise for scalable, high-coherence quantum device fabrication compatible with commercial foundry technologies.

\subsubsection{Layer-by-Layer Process}
\noindent Alternatively, layer-by-layer fabrication approaches define each junction layer through independent lithography steps~\cite{ke2025scaffold, van2024advanced, wang2024}, thereby eliminating the need to deposit and later remove lossy isolation dielectrics. Compared to the traditional tri-layer method, this strategy offers a cleaner integration route that may reduce contamination pathways and, if oxide removal and surface damage are carefully controlled, improve coherence times. While conceptually similar to the evaporation-based techniques described in Subsection~\ref{section:evap}, the key distinction lies in directional thin-film deposition: the junction electrodes are patterned without angled evaporation. Recent work has demonstrated that transmon qubits can be fabricated entirely using foundry-compatible processes, including thin-film deposition and subtractive etching, to define both the top and bottom junction electrodes~\cite{van2024advanced, ke2025scaffold}, as illustrated in Fig.~\ref{cmos:fig:junctionStep}(b,c).

These layer-by-layer methods offer several key advantages. First, depositing the bottom metal layer before lithography enables the use of ultra-high-vacuum thin-film deposition tools while preserving the substrate surface treatments. Second, anisotropic etching of the junction area yields well-defined sidewalls, improving line-edge roughness and minimizing radial variability across the wafer. Third, the tunnel barrier is formed without organic resists, reducing contamination and enhancing junction reliability.

As illustrated in Fig.~\ref{cmos:fig:junctionStep}(b), all-aluminum junction fabrication involves depositing and patterning the Al film to define the bottom electrodes. After ion milling and oxidation to form the $\mathrm{AlO_x}$ barrier, a second Al layer is deposited and patterned to form the top electrodes. Figure~\ref{cmos:fig:junctionStep}(c) shows that the Al JJ bottom electrode and circuit layer are deposited and patterned first. The SiO$_2$ spacer is deposited and patterned to form vias that expose the windows for the JJ and contact areas. The exposed Al bottom layer in the windows is ion-milled and controllably oxidized to form an $\mathrm{AlO_x}$ junction barrier layer, followed by the deposition and patterning of the top Al electrode, which forms the junction. In the end, the SiO$_2$ spacer is removed by a vapor HF process.

A primary challenge of the layer-by-layer approach is the vacuum break between metal deposition steps. Exposure to ambient conditions can lead to the formation of surface oxides on the base electrode, which must be removed before tunnel barrier formation. This is typically addressed via ion milling, in which Ar-ion bombardment strips away native oxides. However, precise calibration is critical: excessive milling can damage the underlying metal, introduce TLSs, and increase surface roughness, while insufficient milling may leave residual oxides that degrade junction performance.

Recent work has demonstrated that wafer-scale, layer-by-layer processes can produce high-coherence devices despite this challenge. For example, foundry-compatible fluxonium qubits fabricated with this method achieved lifetimes $>\mathrm{1~ms}$ although the demonstrated flow still retained evaporation for the top electrode~\cite{wang2024}. Furthermore, the tunnel barrier uniformity achievable through in situ controlled oxidation in these layer-by-layer processes can rival angled evaporation techniques. Nevertheless, compared to the precisely engineered tunnel barrier in tri-layer junctions, this method may exhibit reduced uniformity and reproducibility in critical current density, posing a trade-off between process flexibility and device parameter control. Continued refinement and optimization of process flows will be essential for producing superconducting JJ-based qubits with ultra-high uniformity and coherence.

%% file: sections/63_epitaxial.tex
\subsection{Integration of Novel Junction Materials: From Proof-of-Concept to Scalable Fabrication}
\label{section:epitaxial}

\begingroup
\setlength{\dbltextfloatsep}{8pt}
\setlength{\abovecaptionskip}{4pt}
\setlength{\belowcaptionskip}{0pt}

\begin{figure*}[!t]
    \centering
    \includegraphics[width=17.8cm]{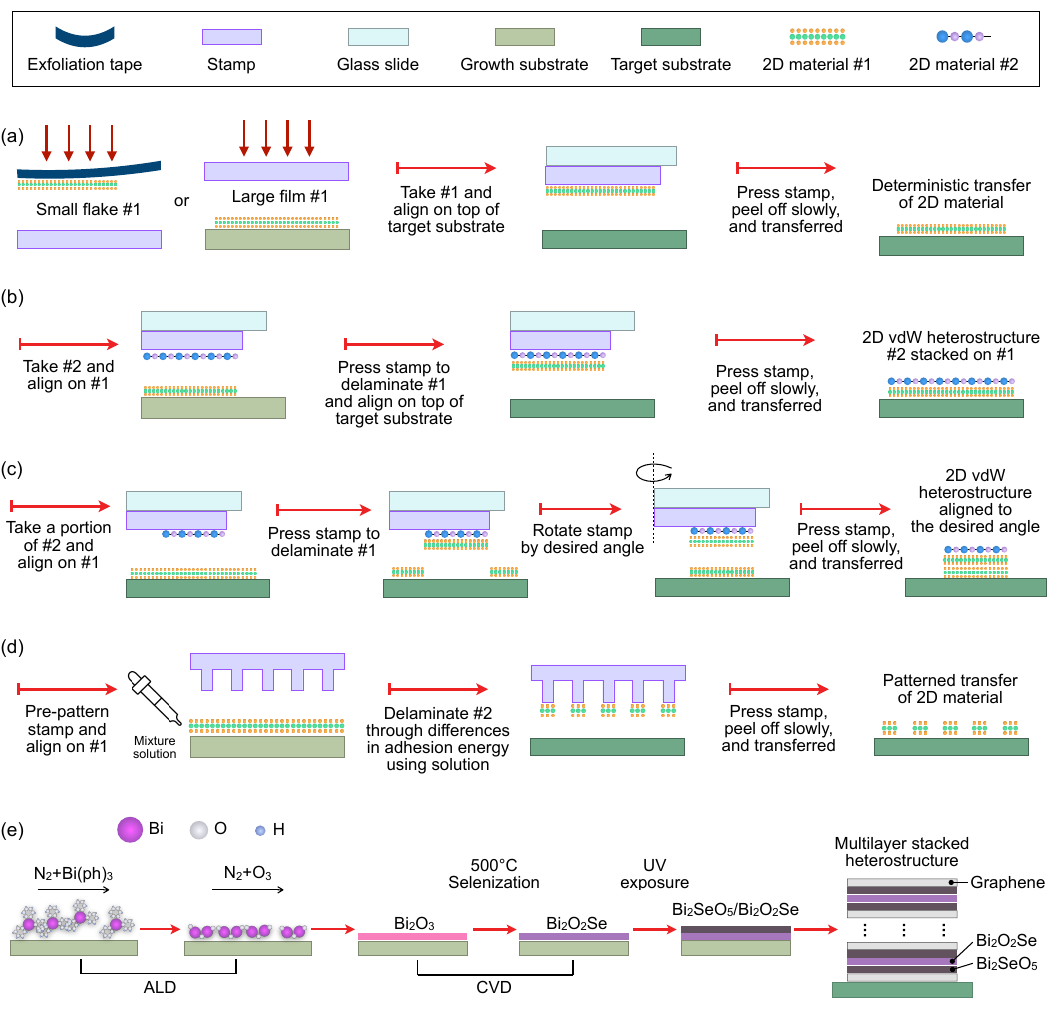}
    \caption{{\textbf{Methods for constructing 2D vdW materials and their heterostructures.} (a) Viscoelastic stamping method for the deterministic transfer of a 2D material~\cite{schranghamer2021review}. (b) Pick-up transfer method for assembling a 2D vdW heterostructure. (c) Tear-and-stack transfer method for a 2D vdW heterostructure aligned to a desired twist angle. (d) Patterned transfer method for large-area patterned transfer of 2D materials. (e) Direct growth and stacking method for multilayer 2D vdW heterostructures.}}
    \label{fig:2Dfab}
\end{figure*}
\endgroup

\noindent Integrating novel junction materials into scalable superconducting circuits requires a staged fabrication pathway. Early proof-of-concept devices often rely on mechanically exfoliated materials or highly controlled epitaxial growth to validate junction physics and interface quality. Long-term adoption in practical quantum hardware, however, demands a transition toward wafer-compatible deposition, patterning, and integration schemes that preserve ultra-low-loss interfaces while delivering high yield and reproducibility. This progression is particularly important for emerging 2D vdW materials, whose promise for superconducting leads, tunnel barriers, ferromagnetic interlayers, and topological weak links is closely tied to the ability to control buried interfaces and structural disorder over large areas.

Conventional SIS JJs have historically been realized using superconductors such as Al~\cite{wu2020vacuo,weides2011coherence,fritz2019optimization}, Nb~\cite{wu2020vacuo}, Ta~\cite{bhatia2024enabling}, Re~\cite{oh2006elimination,kline2011sub}, NbN~\cite{nakamura2011superconducting,wang2013high,qiu2020fabrication,kim2021enhanced}, NiTiN~\cite{supple2024atomic}, and Mg$\mathrm{B_2}$~\cite{li2017high}. Among the available growth routes, molecular beam epitaxy (MBE) provides exceptional control over film purity, stoichiometry, and interface abruptness, but its cost, throughput, and integration complexity limit its widespread use in large-scale fabrication. Homoepitaxial growth of single-crystal superconductors remains challenging, while heteroepitaxy on technologically relevant substrates such as Si(111)~\cite{fritz2019optimization,fritz2019structural} introduces additional constraints related to lattice mismatch, thermal budget, and process integration. As a result, the key challenge is not merely to identify better materials, but to incorporate them into fabrication flows that remain compatible with scalable superconducting circuit manufacturing.

\subsubsection{Conventional Superconductors in CMOS-Compatible Processes}
\noindent CMOS-compatible superconducting process flows on 300~mm wafers increasingly rely on deposition routes such as ALD TaN~\cite{bhatia2024enabling}, sputtered Al, and controlled oxidation on high-resistivity Si substrates~\cite{van2024advanced,verjauw2022path,foroozani2019development,supple2024atomic,peeters2023ultrathin}. These approaches reduce dependence on MBE while still enabling superconducting films and junctions of useful quality. Their appeal lies in process compatibility, wafer-scale throughput, and integration with established semiconductor infrastructure. At the same time, trade-offs remain in film crystallinity, interface abruptness, roughness control, and line-width precision, all of which can directly influence JJ loss, reproducibility, and critical-current spread. Continued refinement of deposition chemistry, surface preparation, and interface engineering will therefore be essential for translating conventional superconductors into commercial foundry-compatible quantum platforms~\cite{nakamura2011superconducting,wang2013high,qiu2020fabrication,kim2021enhanced}.

\subsubsection{Emerging 2D vdW Materials: Current Scalability Limitations}
\noindent In parallel with conventional superconductors, a broad range of 2D vdW materials is being explored for JJs~\cite{wang2024two,farrar2021superconducting,yabuki2016supercurrent,zhao2022josephson,antony2021miniaturizing,wang2022hexagonal,portoles2022tunable,park2022steady,kang2022van,ai2021van,idzuchi2021unconventional,walsh2021josephson,lee2020graphene,kokkoniemi2020bolometer,blaikie2019fast,wu2022field,lin2022zero,kim2024intrinsic,jo2023local,casparis2018superconducting,sagi2024gate,wang2019coherent,qiu2021recent}. Representative families include 2D vdW superconductors such as $\mathrm{NbSe_2}$~\cite{wang2024two,farrar2021superconducting,yabuki2016supercurrent,antony2021miniaturizing,wang2022hexagonal} and $\mathrm{NbS_2}$~\cite{zhao2022josephson}, semimetals such as graphene~\cite{portoles2022tunable,park2022steady,walsh2021josephson,lee2020graphene,kokkoniemi2020bolometer,wang2019coherent,lin2022zero}, topological materials such as $\mathrm{WTe_2}$~\cite{kim2024intrinsic,randle2023gate,zhu2025polarity} and $\mathrm{MoTe_2}$~\cite{chen2023edelstein,zhu2022phase}, and 2D vdW ferromagnets such as $\mathrm{Cr_2Ge_2Te_6}$~\cite{kang2022van,ai2021van,idzuchi2021unconventional}. The main bottleneck is no longer proof-of-concept device physics, but rather reproducible synthesis of low-defect materials and buried interfaces at the wafer scale. Although CVD enables large-area growth, grain boundaries, stoichiometric disorder, and high defect densities often degrade low-temperature electronic and superconducting performance. CVT-based growth can yield higher-quality crystals for research-scale devices~\cite{assouline2019spin,idzuchi2021unconventional,kim2024intrinsic}, but it does not, by itself, address the challenges of large-area integration. As a result, 2D vdW materials will likely enter scalable JJ technologies through a staged pathway: (i) proof-of-concept validation by exfoliation and transfer, (ii) interface verification and engineering, (iii) direct growth and vertical stacking, and finally (iv) foundry-compatible patterning and integration.

\subsubsection{Exfoliation and Transfer}
\noindent Mechanical exfoliation should be viewed primarily as a validation platform rather than a manufacturing solution. It has nonetheless played a central role in early-stage materials screening and device prototyping by enabling researchers to isolate high-quality atomically thin flakes from bulk crystals while minimizing grain boundaries and preserving intrinsic superconducting or correlated-electron properties. This makes exfoliation especially valuable for establishing the relevant figures of merit for candidate JJ materials, including superconducting gap characteristics, interface transparency, transport uniformity, and coherence-related loss channels.

For JJ fabrication, however, the primary challenge is maintaining clean and chemically stable buried interfaces during transfer and assembly. Interfacial contamination, oxidation, trapped bubbles, strain inhomogeneity, and transfer-induced damage can all introduce TLSs, dielectric loss, and spatially nonuniform transport. Considerable effort has therefore been devoted to interface-preserving transfer strategies, including in-vacuum exfoliation~\cite{patil2024pick}, h-BN encapsulation~\cite{Lee2023_hbn_bscc,Martini2023_hbn_bscco,pizzocchero2016hot}, silicon nitride membrane-assisted exfoliation~\cite{wang2023clean}, and inert or trap-free transfer methods~\cite{gant2020system,jung2019reduction}. As summarized in Fig.~\ref{fig:2Dfab}, viscoelastic stamping [Fig.~\ref{fig:2Dfab}(a)] offers a simple deterministic transfer route~\cite{schranghamer2021review}; pick-up transfer [Fig.~\ref{fig:2Dfab}(b)] is widely used for assembling cleaner buried heterointerfaces; tear-and-stack [Fig.~\ref{fig:2Dfab}(c)] enables controlled twist-angle formation; and patterned transfer [Fig.~\ref{fig:2Dfab}(d)] points toward larger-area assembly strategies~\cite{liu2025mass}. Machine-learning-assisted robotic flake identification and transfer may further accelerate prototyping~\cite{masubuchi2018autonomous}. Even so, exfoliation-based JJ fabrication remains fundamentally limited in throughput, reproducibility, and wafer-scale uniformity, and is therefore best regarded as a proof-of-concept route for validating materials and interface concepts before scalable process development.

\subsubsection{Direct Growth and Stack}
\noindent Direct growth becomes compelling only when it can simultaneously address material quality, interface abruptness, and process compatibility. For practical integration into foundry-compatible superconducting circuits, deposition-based methods must eventually replace exfoliation. MBE remains the benchmark for crystalline control and low-defect thin films, but its low throughput and demanding UHV requirements limit large-scale manufacturing. By contrast, ALD offers atomic-scale thickness control and conformality, particularly attractive for tunnel barriers and interface engineering, while sputter- and PVD-based methods offer strong compatibility with established superconducting process flows. CVD enables large-area growth of many 2D vdW materials, although grain boundaries and nonuniform stoichiometry remain major concerns. More recently, hybrid and plasma-assisted routes have emerged as promising bridges between materials quality and process scalability~\cite{moon2025hypotaxy,supple2024atomic,peeters2023ultrathin}.

Figure~\ref{fig:2Dfab}(e) illustrates the broader concept of direct growth and stacking for multilayer 2D vdW heterostructures. In one example, Bi$_2$O$_3$ is grown by ALD and subsequently selenized by CVD to form semiconducting Bi$_2$O$_2$Se, after which UV-assisted intercalative oxidation converts the top layers into an epitaxial Bi$_2$SeO$_5$ dielectric~\cite{tang2025low}. Although this example originates from transistor technology, it demonstrates that wafer-scale vertically integrated vdW heterostacks with tailored interfaces are becoming feasible, which is directly relevant to future JJ tunnel barriers and multilayer superconducting stacks. For quantum devices, the corresponding challenge is to achieve comparable control while simultaneously preserving superconducting properties, minimizing interdiffusion, and suppressing loss associated with defects, traps, local strain, and imperfect buried interfaces.

More generally, scalable 2D vdW JJ fabrication will require high-quality superconducting leads, uniform tunnel barriers or gap interlayers, and, where needed, topological or ferromagnetic proximity layers. The relevant interface engineering must extend beyond SC/SC boundaries to include SC/dielectric, SC/substrate, and SC/vacuum interfaces. Reducing TLS-induced dielectric loss, charge trapping, and local structural disorder is essential for transforming promising 2D vdW junction concepts into reproducible, high-coherence qubits~\cite{de2021materials}. The transition from exfoliation-based all-2D vdW JJs to deposition-based, foundry-compatible processes is therefore not merely a question of throughput, but of establishing a fabrication ecosystem in which materials growth, interface control, patterning, and circuit integration are co-optimized for uniformity, yield, and coherence.

\subsubsection{Post-Fabrication Tunability through MEMS Integration}
\noindent Recent advances in MEMS (microelectromechanical systems) provide a complementary route to improve 2D vdW JJ technologies through post-fabrication trimming and tunability~\cite{yao2021enhanced,lee2024highly,tang2024chip}. MEMS actuators enable controlled adjustment of twist angle, interlayer spacing, and mechanical strain, thereby expanding the design space for highly reproducible quantum devices based on 2D vdW heterostructures or $\mathrm{d/s}$ superconducting bilayers. Such control is particularly attractive for twist-angle-dependent Josephson devices and other twistronic superconducting platforms, where the relevant quantum phenomena can depend sensitively on relative alignment and interfacial spacing~\cite{lin2022zero,naritsuka2025superconductivity}.

Beyond enabling \textit{in situ} tunability, MEMS platforms also provide a practical post-fabrication correction strategy for otherwise unavoidable fabrication errors, including angle misalignment, strain inhomogeneity, and imperfect interfacial contact~\cite{tang2024chip}. Mechanical actuation can additionally suppress wrinkles and trapped bubbles~\cite{nazzari2023reliably}, both of which are known to generate charge traps, dielectric loss, and spatially inhomogeneous transport. In this sense, MEMS integration is not only a route toward new device functionalities but also a potentially valuable tool for improving yield, reproducibility, and interface quality in future twist-angle-sensitive superconducting quantum circuits.

%% file: sections/70_outlook_updated.tex
\section{Outlook}
\label{section:summary}
\noindent The Josephson junction (JJ) has served as the fundamental building block of superconducting quantum circuits for more than six decades and remains central to modern quantum technology~\cite{bravyi2022future,mohseni2024build,devoret2013superconducting}. What has changed is the scale at which it must now operate. A junction that performs adequately in a two-qubit experiment may be insufficient in a processor containing thousands of qubits, where fabrication variability, frequency crowding, and device imperfections accumulate across the chip~\cite{bravyi2022future,acharya2025}. In such systems, the worst-performing junction can ultimately limit the entire processor's performance. The central question for the field is therefore no longer whether JJs can support quantum computation, but whether they can do so with the uniformity, stability, and scalability required for fault-tolerant architectures~\cite{blais2021circuit,acharya2025}.

In Section~\ref{section:computing}, we examined five key challenges that will shape the future of JJs in quantum processors: yield and reproducibility, dissipation, tunability, device footprint, and intrinsic noise protection. On the conventional Al/AlO$_x$/Al platform, sustained engineering progress has driven coherence times beyond one millisecond and junction uniformity to the few-percent level~\cite{bland2025millisecond,place2021new}. These advances have enabled rapid progress in superconducting quantum processors, including multi-qubit devices and early demonstrations of logical error suppression~\cite{bravyi2022future,Arute2019supremacy,acharya2025}. At the same time, several limitations are becoming increasingly apparent as systems scale. In particular, the amorphous nature of the AlO$_x$ tunnel barrier introduces stochastic TLSs that contribute to energy loss and frequency instability~\cite{muller2019towards,lubchenko2007microscopic,ganjam2024surpassing}. Although extensive process optimization has substantially reduced these effects, their microscopic origins suggest that further gains may ultimately require either alternative barrier materials or new qubit architectures designed to mitigate these loss mechanisms.

\begin{table*}[b]
\centering
\caption{Coherence benchmarks across JJ material platforms discussed.}
\label{tab:coherence}
\begin{tabular}{l c c l c c}
\hline\hline
Platform & Best $T_1$ & Best $T_2$ & Junction Type & Year & Ref. \\
\hline
Al/AlO$_x$/Al (mergemon) JJ        & ${\sim}\,100$~\textmu s & ${\sim}\,50$~\textmu s & Thick tunnel barrier & 2020 & \cite{zhao2020merged,mamin2021merged} \\
Nb/Al-AlO$_x$/Nb (trilayer) JJ     & ${\sim}\,60$~\textmu s  & ${\sim}\,35$~\textmu s              & Conventional tunnel barrier              & 2024    & \cite{anferov2024improved}
\\
Al/AlO$_x$/Al (layer-by-layer fluxonium) JJ  & ${\sim}\,1$~ms       & ${\sim}\,0.9$~ms    & Conventional tunnel barrier              & 2025    & \cite{wang2024}\\
Al/AlO$_x$/Al (Ta shunt) JJ        & ${\sim}\,1.6$~ms    & ${\sim}\,1.2$~ms              & Conventional tunnel barrier              & 2025    & \cite{bland2025millisecond}
\\
Al/InGaAs/InAs/InGaAs/Al gatemon qubits                    & ${\sim}\,1$~\textmu s   & ${\sim}\,2$~\textmu s  & SC/SM-based 2DEG system             & 2018    & \cite{casparis2018superconducting}
\\
Al/h-BN/graphene/h-BN/Al gatemon qubits                & ${\sim}\,30$~ns   & ${\sim}\,50$~ns & 2D vdW heterostructure               & 2019    & \cite{wang2019coherent}
\\
Al/AlO$_x$/Al (NbSe$_2$/h-BN/NbSe$_2$ shunt) SQUID           & ${\sim}\,50$~\textmu s    & ${\sim}\,70$~\textmu s   & Conventional + 2D vdW             & 2022    & \cite{wang2022hexagonal}
\\
NbSe$_2$/WSe$_2$/NbSe$_2$ JJ-based mergemon qubits              & ${\sim}\,2$~\textmu s   & ${\sim}\,4$~\textmu s    & 2D vdW heterostructure       & 2025    & \cite{Balgley2025coherent}
\\
\hline\hline
\end{tabular}
\end{table*}

Several emerging material platforms offer promising pathways to address these challenges. Crystalline tunnel barriers, SC/SM junction-based 2DEG systems, 2D vdW materials and their heterostructures, and ferromagnetic interlayers each target different limitations of conventional devices. Crystalline and epitaxial barriers aim to reduce structural disorder and TLS density, while semiconducting materials enable electrostatic tunability that can mitigate frequency collisions and support new circuit functionalities~\cite{casparis2018superconducting,kjaergaard2017transparent}. 2D vdW materials and their heterostructures can provide cleaner assembled interfaces with reduced dangling-bond disorder compared with conventional amorphous barriers, while their atomically thin nature also enables device miniaturization~\cite{novoselov20162d,lee2019two,antony2021miniaturizing}. At the same time, unconventional superconducting and magnetic systems introduce the possibility of intrinsically protected qubit architectures based on higher-order tunneling processes, such as Cooper-quartet tunneling~\cite{patel2024d,brosco2024superconducting}. Despite these promising directions, most emerging platforms remain at an early stage. Coherence times reported for many emerging junction platforms remain below those of the most mature aluminum-based qubits~\cite{DElia2025coherent,Balgley2025coherent} (Table~\ref{tab:coherence}). However, it is worth recalling that aluminum-based qubits themselves improved by a factor of $10^{6}$ over roughly two decades, from nanoseconds in the early 2000s to milliseconds today, and that much of this improvement arose not from the junction alone, but from advances in substrates, packaging, filtering, and circuit design once the underlying junction platform was established. Note that unconventional superconducting qubits can have much longer lifetimes~\cite{pop2014coherent,lin2018demonstration,earnest2018realization, gyenis2021experimental, hassani2023inductively,nguyen2025superconducting}. A similar trajectory may be possible for emerging platforms, provided that materials quality, interface control, and scalable fabrication improve in parallel.

In parallel with materials innovation, the fabrication advances discussed in Section~\ref{section:fab} will be essential for the next generation of quantum processors. Foundry-compatible JJ technologies based on optical lithography, trilayer deposition, and etch-defined junctions have recently demonstrated coherence approaching that of traditional shadow-evaporated devices, while also offering improved prospects for wafer-scale uniformity~\cite{tolpygo2014fabrication,verjauw2022path}. These developments represent an important step toward industrial-scale quantum integrated circuits. A useful long-term analogy is the fabless, foundry model of the semiconductor industry, in which standardized process design rules decouple circuit innovation from manufacturing scale. Superconducting quantum technology has not yet reached this level of maturity, but early signs of a similar ecosystem are emerging. Establishing reproducible junction processes, well-defined fabrication tolerances, and shared process design rules will therefore be essential for enabling scalable quantum hardware development across both academic and industrial platforms. Of equal importance is the development of standardized junction metrology, analogous to transistor compact modeling in the semiconductor industry, that can systematically relate barrier microstructure and transport properties to qubit-level performance, thereby enabling rapid iteration across fabrication processes and material platforms.

For emerging materials platforms, fabrication presents a different challenge. Many promising demonstrations, particularly in 2D vdW systems, currently rely on mechanical exfoliation and manual assembly~\cite{Geim2013Van}. While these approaches are invaluable for proof-of-concept experiments and materials validation, they are inherently difficult to scale to large multi-qubit processors. Bridging this gap will require advances in wafer-scale materials synthesis, deterministic material placement, and hybrid integration strategies that combine conventional superconducting circuitry with the localized incorporation of new materials only where their unique functionality is required.

Looking ahead, several milestones would mark major inflection points for the field. First, demonstrating qubits based on crystalline materials or 2D vdW JJs with coherence times approaching $T_1 > 100~\mu$s would establish that non-amorphous barriers can support coherence well beyond the proof-of-concept regime~\cite{bland2025millisecond,place2021new}. Second, realizing a qubit in which higher-order tunneling processes, such as intrinsic Cooper-quartet tunneling, dominate the Josephson energy would provide a new route toward hardware-level noise protection without relying on complex circuit elements~\cite{patel2024d,brosco2024superconducting}. Third, foundry-fabricated JJs that simultaneously achieve high coherence, tight parameter control, and high yield across large wafers would mark a decisive step toward scalable quantum integrated circuits~\cite{tolpygo2014fabrication,van2024advanced}.

The remarkable progress of superconducting quantum processors over the past decade, from few-qubit experiments to processors containing hundreds or thousands of qubits, has been built on the JJ~\cite{bravyi2022future}. As quantum processors continue to scale, the junction itself cannot be treated as a mature, static component. Its materials, fabrication processes, and circuit roles must evolve together with the demands of practical quantum computing. The developments surveyed in this review suggest that the next era of superconducting quantum hardware will be shaped not only by scaling up larger circuits, but also by better junctions—designed, fabricated, and integrated with the same rigor that transformed semiconductor electronics.

%% file: sections/80_acknowledgements.tex
\section*{ACKNOWLEDGMENTS}
\noindent
The authors thank Kan-Heng Lee, Larry Chen, Haleem Kim, Hanho Lee, and William Strickland for fruitful discussions. H.K. and L.B.N. are grateful to Christopher Spitzer for his valuable assistance. This work was supported by the National Research Foundation of Korea (NRF) Grant, funded by the Korean Government's Ministry of Science and ICT (MSIT) with Grant Nos.~RS-2022-NR068223, RS-2023-NR068116, RS-2024-00333664, RS-2024-00352458, RS-2024-00353348, and RS-2025-00561110; the Quantum Systems Accelerator (QSA) Grant of the National Quantum Information Science Research Centers, funded by the U.S. Department of Energy's Office of Science; and the Global Value-Up 10X Project grant funded by the Gwangju Institute of Science and Technology (GIST) in 2026.

%% file: sections/81_abbre.tex
\newpage
\noindent
\appendix 
\section*{\MakeUppercase{Appendix A: List of Abbreviations}}
\dashconnect{ABS}{Andreev Bound States}
\dashconnect{ALD}{Atomic Layer Deposition}
\dashconnect{BOE}{Buffered Oxide Etchant}
\dashconnect{CGT}{Cr$_2$Ge$_2$Te$_6$}
\dashconnect{CMOS}{Complementary Metal-Oxide-Semiconductor}
\dashconnect{CPR}{Current-Phase Relation}
\dashconnect{cQED}{Circuit Quantum Electrodynamics}
\dashconnect{CQT}{Cooper-Quartet Tunneling}
\dashconnect{CVD}{Chemical Vapor Deposition}
\dashconnect{CVT}{Chemical Vapor Transport}
\dashconnect{DUV}{Deep Ultraviolet}
\dashconnect{EBL}{Electron-Beam Lithography}
\dashconnect{EOT}{Effective Oxide Thickness}
\dashconnect{FET}{Field-Effect Transistor}
\dashconnect{IC}{Integrated Circuit}
\dashconnect{JJ}{Josephson Junction}
\dashconnect{MBE}{Molecular Beam Epitaxy}
\dashconnect{MEMS}{Microelectromechanical Systems}
\dashconnect{NISQ}{Noisy Intermediate-Scale Quantum}
\dashconnect{PVD}{Physical Vapor Deposition}
\dashconnect{R\&D}{Research and Development}
\dashconnect{SC}{Superconductor}
\dashconnect{SM}{Semiconductor}
\dashconnect{SQUID}{Superconducting Quantum Interference Device}
\dashconnect{TEM}{Tunneling Electron Microscopy}
\dashconnect{TRS}{Time-Reversal Symmetry}
\dashconnect{TSV}{Through-Silicon Vias}
\dashconnect{TLS}{Two-Level System}
\dashconnect{VHF}{Vapor Hydrofluoric Acid}
\dashconnect{vdW}{van der Waals}
\dashconnect{2D}{Two-Dimensional}
\dashconnect{2DEG}{Two-Dimensional Electron Gas}

\newpage
\section*{\MakeUppercase{Appendix B: Current-phase harmonics}}

For a short superconducting weak link with transmission eigenvalues $\{\tau_i\}$, the zero-temperature CPR can be written as
\begin{equation}
I(\varphi)
= \sum_i \frac{e\Delta}{\hbar}\,
\frac{\tau_i \sin\varphi}
{2\sqrt{1 - \tau_i \sin^2(\varphi/2)}} .
\label{eq:CPR_exact}
\end{equation}

Using the relation $\sin\varphi = 2\sin(\varphi/2)\cos(\varphi/2)$, this becomes
\begin{equation}
I_i(\varphi)
= \frac{e\Delta}{\hbar}\,
\tau_i \,
\sin\frac{\varphi}{2}\cos\frac{\varphi}{2}
\left[1 - \tau_i \sin^2\frac{\varphi}{2}\right]^{-1/2}.
\end{equation}

The square root can be expanded using the binomial series with $x=\tau_i\sin^2(\varphi/2)$.
\begin{equation}
(1-x)^{-1/2}
= \sum_{n=0}^{\infty}
\binom{-\tfrac12}{n} (-x)^n
= \sum_{n=0}^{\infty}
\frac{(2n)!}{2^{2n}(n!)^2}\, x^n ,
\end{equation}

Substituting, we obtain
\begin{equation}
I_i(\varphi)
= \frac{e\Delta}{\hbar}
\sum_{n=0}^{\infty}
\frac{(2n)!}{2^{2n}(n!)^2}\,
\tau_i^{\,n+1}
\sin^{2n+1}\frac{\varphi}{2}
\cos\frac{\varphi}{2}.
\label{eq:series_intermediate}
\end{equation}

Each term $\sin^{2n+1}(\varphi/2)\cos(\varphi/2)$ can be expressed as a finite Fourier sine series, where $a_{n,k}$ are numerical coefficients.
\begin{equation}
\sin^{2n+1}\theta\,\cos\theta
= \sum_{k=1}^{n+1} a_{n,k}\,\sin(2k\theta),
\end{equation}

Setting $\theta=\varphi/2$ shows that
Eq.~\eqref{eq:series_intermediate} contains harmonics $\sin(k\varphi)$ with
$k\le n+1$. Therefore, the current admits a Fourier expansion with $I_k=\sum_i I_{k i}$.
\begin{equation}
I_i(\varphi)
= \sum_{k=1}^{\infty} I_{k i}\,\sin(k\varphi),
\qquad
I(\varphi)=\sum_i I_i(\varphi)=\sum_{k} I_k \sin(k\varphi),
\end{equation}

\subsubsection*{Small-transparency expansion}

For $\tau_i\ll1$, expanding Eq.~\eqref{eq:CPR_exact} to third order yields
\begin{align}
I_i(\varphi)
\simeq \frac{e\Delta}{\hbar}\Big[
&\left(\frac{\tau_i}{2}
      +\frac{\tau_i^2}{16}
      +\frac{15\tau_i^3}{512}\right)\sin\varphi
-\left(\frac{\tau_i^2}{32}
      +\frac{3\tau_i^3}{128}\right)\sin(2\varphi)
\nonumber\\
&\hspace{2cm}
+\frac{3\tau_i^3}{512}\sin(3\varphi)
+\mathcal{O}(\tau_i^4)
\Big].
\label{eq:small_tau_CPR}
\end{align}

This explicitly demonstrates that higher harmonics scale as
$I_{k i}\propto \tau_i^{k}$ in the tunneling limit, while increasing transmission
leads to a strongly non-sinusoidal current--phase relation.
\newpage

%% file: sections/82_biography.tex
\pagebreak
\newcommand{\biobox}[2]{%
\noindent
\begin{minipage}[t]{0.18\textwidth}
\vspace{0pt}
\includegraphics[width=\linewidth]{#1}
\end{minipage}\hspace{0.04\textwidth}%
\begin{minipage}[t]{0.78\textwidth}
\vspace{0pt}
#2
\end{minipage}

\vspace{1.2em}
}

\biobox{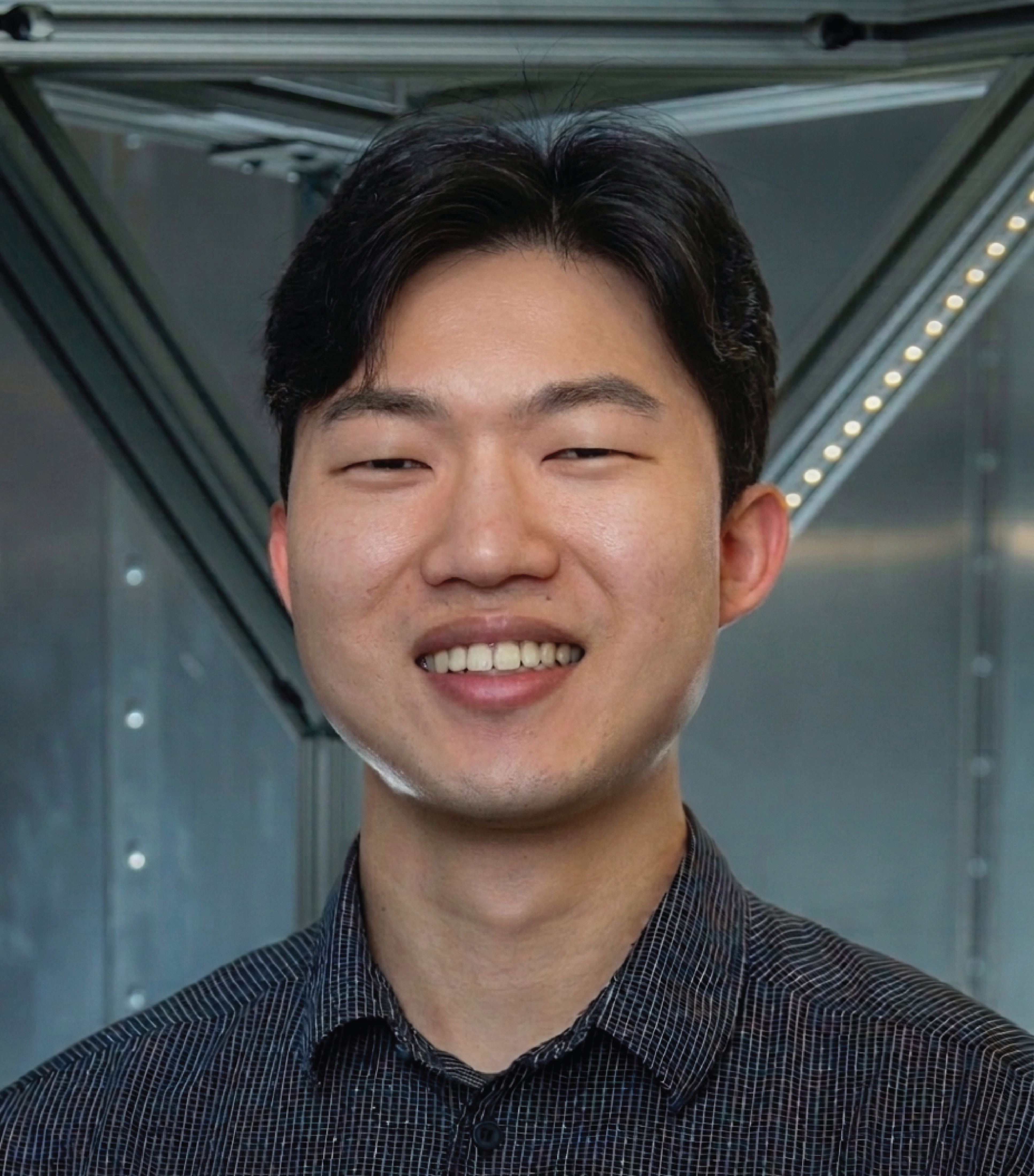}{
\textbf{Hyunseong "Linus" Kim} is a Ph.D. candidate in the Department of Physics at the University of California, Berkeley. He received his B.S. in Physics from the California Institute of Technology. During his Ph.D., he led the development of the gridium qubit, a novel superconducting qubit architecture featuring intrinsic noise protection. In recognition of this work, he was awarded the Kavli ENSI Graduate Student Fellowship, under which he is advancing the theoretical framework and experimental control techniques for the gridium. His previous research includes investigating laser annealing of Josephson junctions to enable component-level frequency targeting in superconducting quantum processors, as well as mitigating dark counts in superconducting nanowire single-photon detectors (SNSPDs). His broader research interests encompass Hamiltonian engineering for noise-protected superconducting circuits, mesoscopic Josephson physics, and superconducting circuit design.
}

\biobox{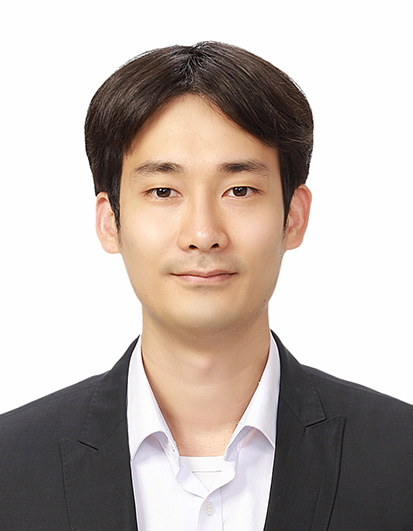}{
\textbf{Gyunghyun Jang} received his B.S. degree in Physics from Chonnam National University in 2022, where he was involved in spintronics experiments during his senior year. He joined the Department of Semiconductor Engineering at the Gwangju Institute of Science and Technology (GIST) in the same year to pursue his M.S.–Ph.D. studies. His research focuses on superconducting transmon qubits based on two-dimensional van der Waals materials. His broader interests include superconducting quantum devices and Josephson junction–based systems, such as Josephson diodes, bolometers, and parametric amplifiers, as well as emerging electronic and optoelectronic devices.
}

\biobox{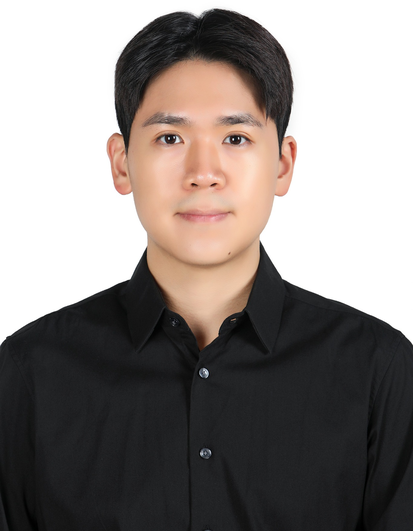}{
\textbf{Seungwon Jin} received his B.Sc. in Physics from Sogang University, and he conducted quantum experiments on nitrogen-vacancy centers during his senior year. He subsequently worked on superconducting quantum systems at Korea University, focusing on the development of a 3D bosonic platform. He is currently pursuing graduate studies at the National University of Singapore, where he continues his research on 3D bosonic circuit QED devices with a focus on light--matter interactions and multimode dynamics. His research interests lie in continuous-variable encoding with bosonic modes, novel quantum control techniques, and the investigation of physical phenomena using circuit QED systems.
}

\biobox{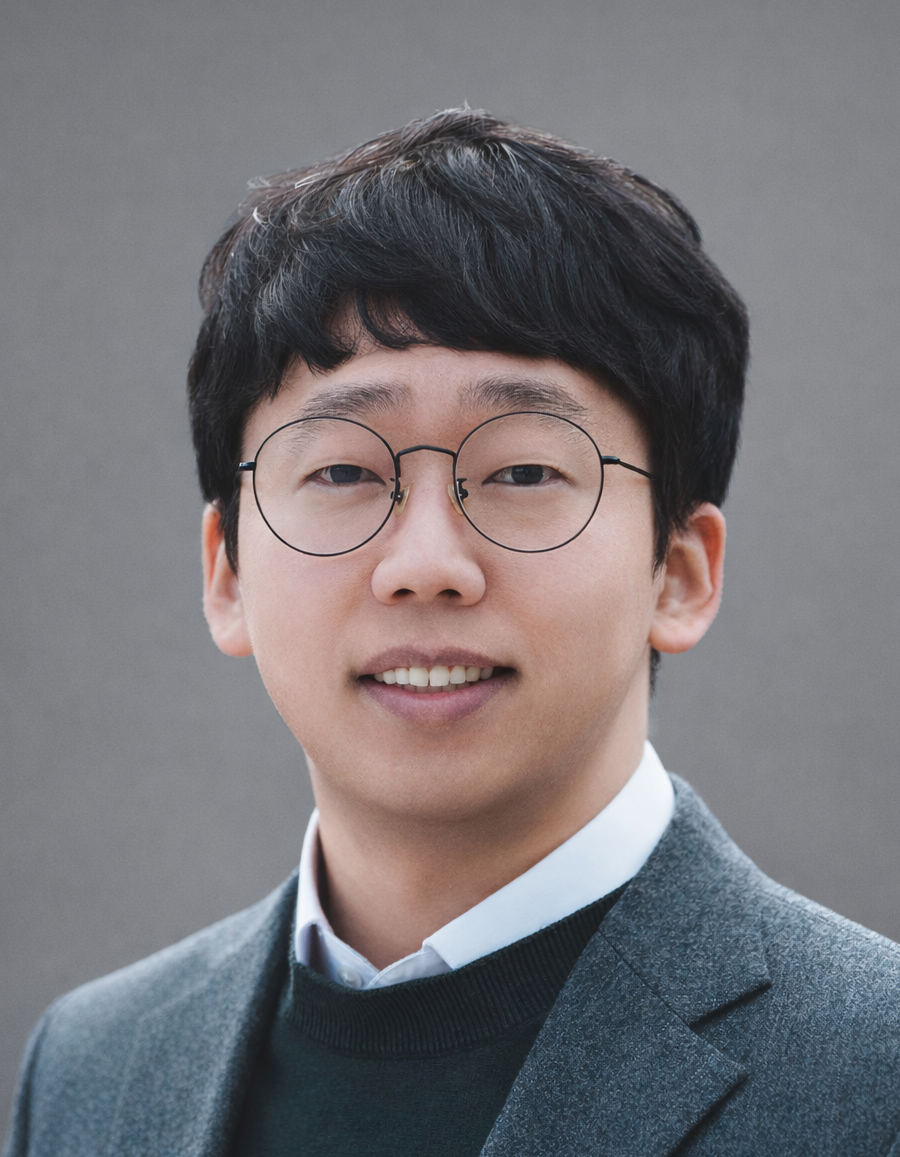}{
\textbf{Yosep Kim} is an Assistant Professor in the Department of Physics at Korea University. He received his B.S. in Electrical and Computer Engineering from the Ulsan National Institute of Science and Technology (UNIST), graduating summa cum laude, and his Ph.D. in Physics from Pohang University of Science and Technology (POSTECH). His doctoral research focused on photonic quantum computing and generalized quantum measurements. After completing his Ph.D., he worked as a postdoctoral researcher at Lawrence Berkeley National Laboratory, where he studied superconducting quantum circuits and quantum processors. He later joined the Korea Institute of Science and Technology (KIST) as a Senior Researcher before starting his current position at Korea University. His research focuses on quantum information science across photonic and superconducting platforms. He has contributed to the development of high-fidelity multi-qubit gates and programmable interactions in Floquet qubits for superconducting quantum processors. 
}

\biobox{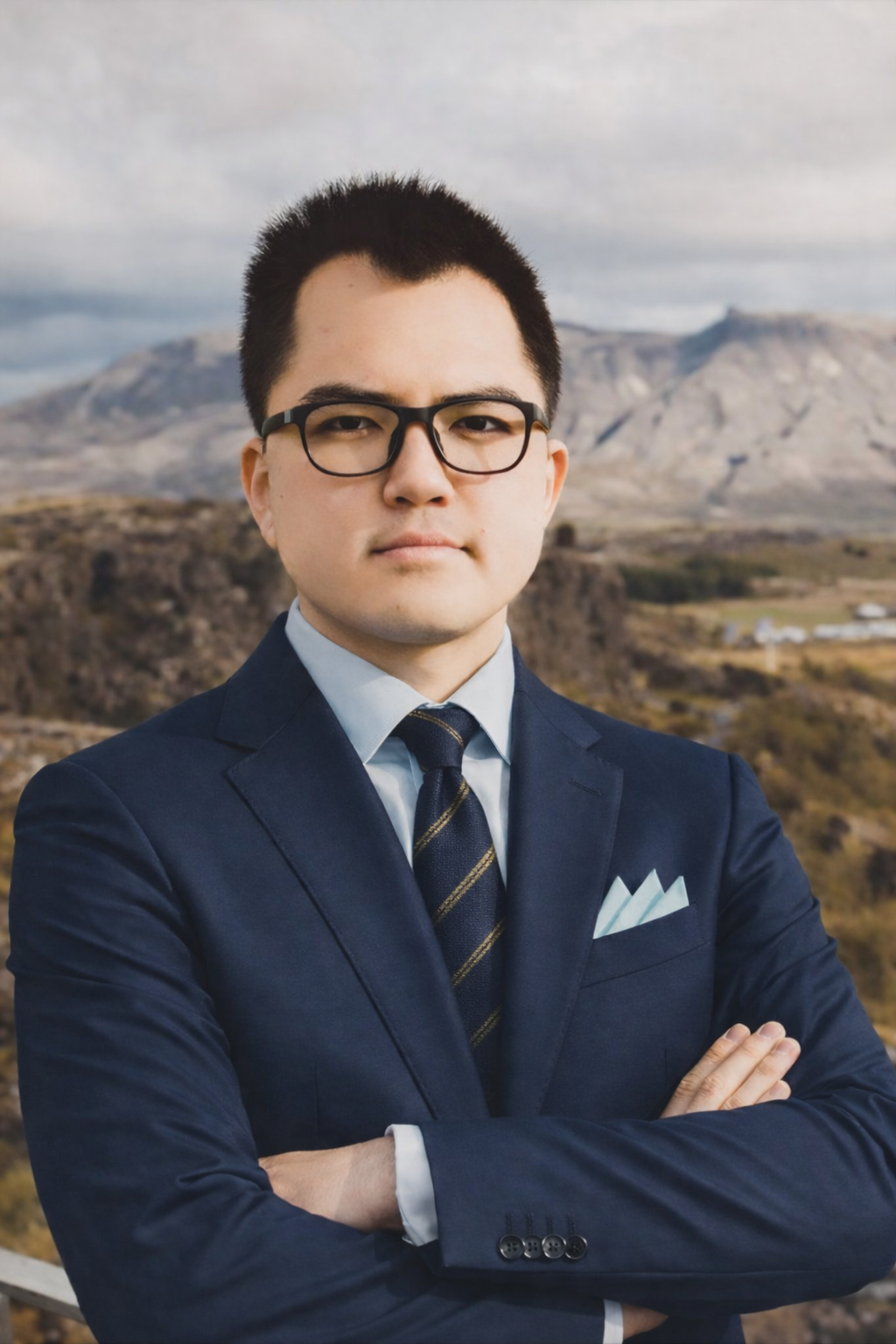}{
\textbf{Long Bao Nguyen} is currently a Quantum Applied Scientist at the AWS Center for Quantum Computing, where he develops advanced protocols to optimize the performance of quantum architectures based on cat qubits. Previously, as a quantum scientist at the University of California, Berkeley, and Lawrence Berkeley National Laboratory, he pioneered the application of Floquet engineering to tailor interactions between qudits and introduced the superconducting gridium qubit. During his Ph.D. at the University of Maryland, College Park, he discovered long coherence times in fluxonium qubits and achieved the first entanglement between multiple fluxonium qubits. These achievements earned him the Condensed Matter Graduate Fellowship, the Northrop Grumman Fellowship, and a Distinguished Dissertation Award. He holds a B.Sc. from the University of Texas at Dallas, where he graduated Summa Cum Laude with a first-place senior design project award. His current research interests span from mesoscopic Josephson physics to hardware-efficient fault-tolerant quantum information processing.
}

\biobox{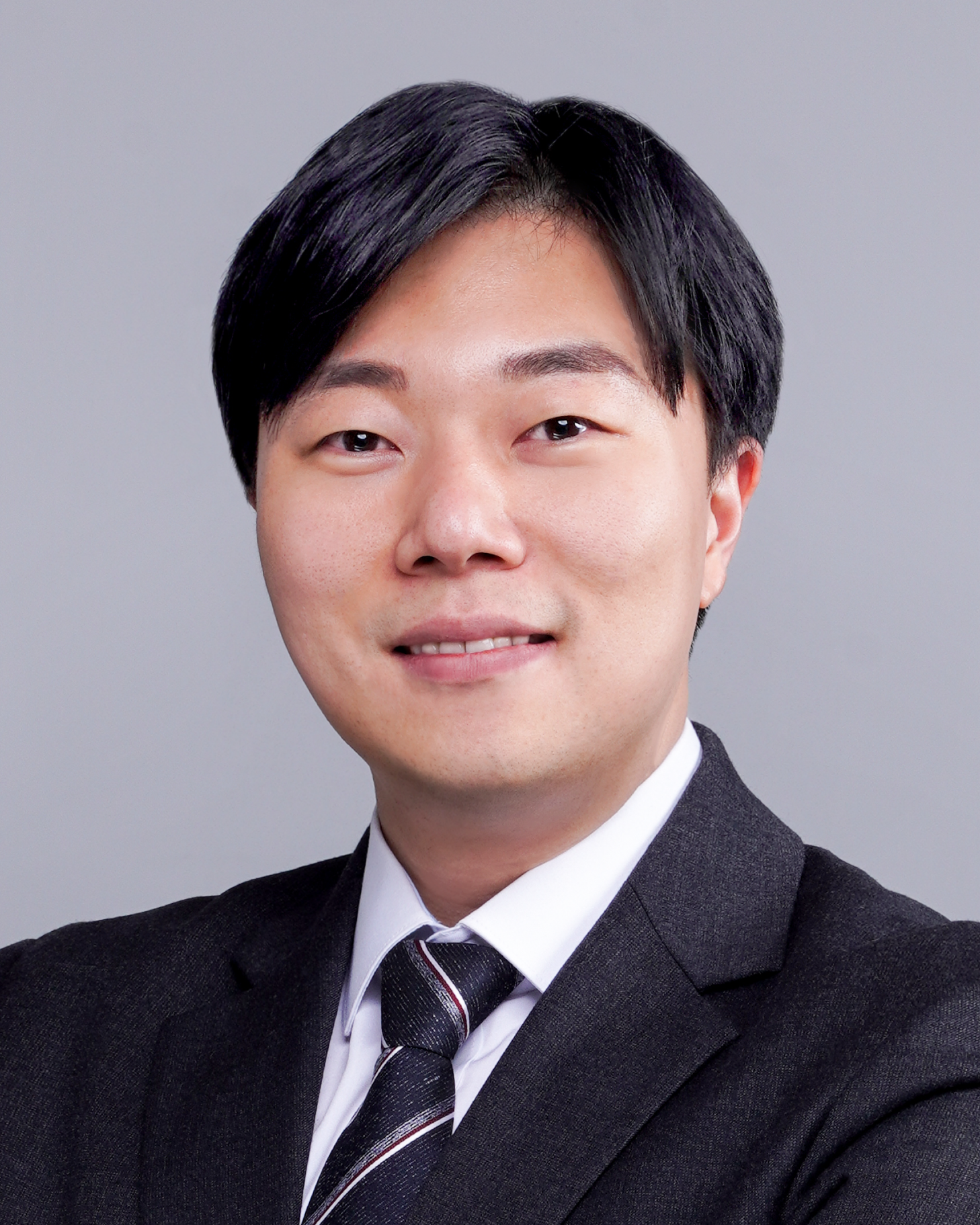}{
\textbf{Hoon Hahn Yoon} double majored in Device Physics and Mechanical System Design and Manufacturing for his B.S. degree from the Ulsan National Institute of Science and Technology and delivered the graduation speech as a valedictorian at the commencement ceremony in 2014. He received his Ph.D. in Physics from the same university in 2020. He worked as an Intern Researcher at the Single-Electron Quantum Device Team, Quantum Technology Institute, Korea Research Institute of Standards and Science in 2012, a Postdoctoral Researcher at the Department of Physics, Ulsan National Institute of Science and Technology in 2020, a Postdoctoral Researcher at the Department of Electronics and Nanoengineering, Aalto University from 2020 to 2022, and a Staff Engineer at the Device Part, Process Architecture Team, Foundry Business Department, Samsung Electronics from 2022 to 2023. He then joined the Department of Semiconductor Engineering (Full) and Department of Electrical Engineering and Computer Science (Adjunct) at the Gwangju Institute of Science and Technology in 2023 as an Assistant Professor. He is a member of the steering committee of the GIST Advanced AI Semiconductor Fab Center (ANGELS: Advanced Nanofab at GIST for Emerging Low-power Semiconductors). He was awarded the Global Ph.D. Fellowship from the National Research Foundation of Korea in 2015, the Academy Research Fellowship from the Academy of Finland in 2022, and the 2DM Emerging Young Scientists Award from the journal \textit{2D Materials} by IOP Publishing in 2025. His research interests are focused on two-dimensional materials for artificial intelligence semiconductor systems and quantum computing systems, including artificial intelligence sensors, more-than-Moore devices, electron quantum optics, and quantum phase transitions.}